\documentclass[11pt]{article}
\usepackage{tcolorbox}
\usepackage[T1]{fontenc}
\usepackage{mathpazo}
\usepackage{graphicx}
\usepackage{dirtytalk}
\usepackage{bbold}
\usepackage{caption}
\DeclareCaptionLabelFormat{nolabel}{}
\usepackage{chngcntr} 
\usepackage{enumerate} 
\usepackage{geometry} 
\usepackage{amsmath} 
\usepackage{amssymb} 
\usepackage{textcomp}
    \AtBeginDocument{%
        
    }
\usepackage{upquote} 
\usepackage{eurosym} 
\usepackage[mathletters]{ucs} 
\usepackage[utf8x]{inputenc} 
\usepackage{fancyvrb} 
\usepackage{grffile} 
\usepackage{longtable} 
\usepackage{booktabs} 
\usepackage[inline]{enumitem} 
\usepackage[normalem]{ulem} 

\usepackage[pdftex,pagebackref,colorlinks,linkcolor=blue,filecolor = blue, citecolor = magenta, urlcolor  = JungleGreen]{hyperref}

    \definecolor{urlcolor}{rgb}{0,.145,.698}
    \definecolor{linkcolor}{rgb}{.71,0.21,0.01}
    \definecolor{citecolor}{rgb}{.12,.54,.11}

    \definecolor{ansi-black}{HTML}{3E424D}
    \definecolor{ansi-black-intense}{HTML}{282C36}
    \definecolor{ansi-red}{HTML}{E75C58}
    \definecolor{ansi-red-intense}{HTML}{B22B31}
    \definecolor{ansi-green}{HTML}{00A250}
    \definecolor{ansi-green-intense}{HTML}{007427}
    \definecolor{ansi-yellow}{HTML}{DDB62B}
    \definecolor{ansi-yellow-intense}{HTML}{B27D12}
    \definecolor{ansi-blue}{HTML}{208FFB}
    \definecolor{ansi-blue-intense}{HTML}{0065CA}
    \definecolor{ansi-magenta}{HTML}{D160C4}
    \definecolor{ansi-magenta-intense}{HTML}{A03196}
    \definecolor{ansi-cyan}{HTML}{60C6C8}
    \definecolor{ansi-cyan-intense}{HTML}{258F8F}
    \definecolor{ansi-white}{HTML}{C5C1B4}
    \definecolor{ansi-white-intense}{HTML}{A1A6B2}

    \DefineVerbatimEnvironment{Highlighting}{Verbatim}{commandchars=\\\{\}}

    \title{Introduction to python-igraph}

\makeatletter
\def\PY@reset{\let\PY@it=\relax \let\PY@bf=\relax%
    \let\PY@ul=\relax \let\PY@tc=\relax%
    \let\PY@bc=\relax \let\PY@ff=\relax}
\def\PY@tok#1{\csname PY@tok@#1\endcsname}
\def\PY@toks#1+{\ifx\relax#1\empty\else%
    \PY@tok{#1}\expandafter\PY@toks\fi}
\def\PY@do#1{\PY@bc{\PY@tc{\PY@ul{%
    \PY@it{\PY@bf{\PY@ff{#1}}}}}}}
\def\PY#1#2{\PY@reset\PY@toks#1+\relax+\PY@do{#2}}

\expandafter\def\csname PY@tok@gd\endcsname{\def\PY@tc##1{\textcolor[rgb]{0.63,0.00,0.00}{##1}}}
\expandafter\def\csname PY@tok@gu\endcsname{\let\PY@bf=\textbf\def\PY@tc##1{\textcolor[rgb]{0.50,0.00,0.50}{##1}}}
\expandafter\def\csname PY@tok@gt\endcsname{\def\PY@tc##1{\textcolor[rgb]{0.00,0.27,0.87}{##1}}}
\expandafter\def\csname PY@tok@gs\endcsname{\let\PY@bf=\textbf}
\expandafter\def\csname PY@tok@gr\endcsname{\def\PY@tc##1{\textcolor[rgb]{1.00,0.00,0.00}{##1}}}
\expandafter\def\csname PY@tok@cm\endcsname{\let\PY@it=\textit\def\PY@tc##1{\textcolor[rgb]{0.25,0.50,0.50}{##1}}}
\expandafter\def\csname PY@tok@vg\endcsname{\def\PY@tc##1{\textcolor[rgb]{0.10,0.09,0.49}{##1}}}
\expandafter\def\csname PY@tok@vi\endcsname{\def\PY@tc##1{\textcolor[rgb]{0.10,0.09,0.49}{##1}}}
\expandafter\def\csname PY@tok@vm\endcsname{\def\PY@tc##1{\textcolor[rgb]{0.10,0.09,0.49}{##1}}}
\expandafter\def\csname PY@tok@mh\endcsname{\def\PY@tc##1{\textcolor[rgb]{0.40,0.40,0.40}{##1}}}
\expandafter\def\csname PY@tok@cs\endcsname{\let\PY@it=\textit\def\PY@tc##1{\textcolor[rgb]{0.25,0.50,0.50}{##1}}}
\expandafter\def\csname PY@tok@ge\endcsname{\let\PY@it=\textit}
\expandafter\def\csname PY@tok@vc\endcsname{\def\PY@tc##1{\textcolor[rgb]{0.10,0.09,0.49}{##1}}}
\expandafter\def\csname PY@tok@il\endcsname{\def\PY@tc##1{\textcolor[rgb]{0.40,0.40,0.40}{##1}}}
\expandafter\def\csname PY@tok@go\endcsname{\def\PY@tc##1{\textcolor[rgb]{0.53,0.53,0.53}{##1}}}
\expandafter\def\csname PY@tok@cp\endcsname{\def\PY@tc##1{\textcolor[rgb]{0.74,0.48,0.00}{##1}}}
\expandafter\def\csname PY@tok@gi\endcsname{\def\PY@tc##1{\textcolor[rgb]{0.00,0.63,0.00}{##1}}}
\expandafter\def\csname PY@tok@gh\endcsname{\let\PY@bf=\textbf\def\PY@tc##1{\textcolor[rgb]{0.00,0.00,0.50}{##1}}}
\expandafter\def\csname PY@tok@ni\endcsname{\let\PY@bf=\textbf\def\PY@tc##1{\textcolor[rgb]{0.60,0.60,0.60}{##1}}}
\expandafter\def\csname PY@tok@nl\endcsname{\def\PY@tc##1{\textcolor[rgb]{0.63,0.63,0.00}{##1}}}
\expandafter\def\csname PY@tok@nn\endcsname{\let\PY@bf=\textbf\def\PY@tc##1{\textcolor[rgb]{0.00,0.00,1.00}{##1}}}
\expandafter\def\csname PY@tok@no\endcsname{\def\PY@tc##1{\textcolor[rgb]{0.53,0.00,0.00}{##1}}}
\expandafter\def\csname PY@tok@na\endcsname{\def\PY@tc##1{\textcolor[rgb]{0.49,0.56,0.16}{##1}}}
\expandafter\def\csname PY@tok@nb\endcsname{\def\PY@tc##1{\textcolor[rgb]{0.00,0.50,0.00}{##1}}}
\expandafter\def\csname PY@tok@nc\endcsname{\let\PY@bf=\textbf\def\PY@tc##1{\textcolor[rgb]{0.00,0.00,1.00}{##1}}}
\expandafter\def\csname PY@tok@nd\endcsname{\def\PY@tc##1{\textcolor[rgb]{0.67,0.13,1.00}{##1}}}
\expandafter\def\csname PY@tok@ne\endcsname{\let\PY@bf=\textbf\def\PY@tc##1{\textcolor[rgb]{0.82,0.25,0.23}{##1}}}
\expandafter\def\csname PY@tok@nf\endcsname{\def\PY@tc##1{\textcolor[rgb]{0.00,0.00,1.00}{##1}}}
\expandafter\def\csname PY@tok@si\endcsname{\let\PY@bf=\textbf\def\PY@tc##1{\textcolor[rgb]{0.73,0.40,0.53}{##1}}}
\expandafter\def\csname PY@tok@s2\endcsname{\def\PY@tc##1{\textcolor[rgb]{0.73,0.13,0.13}{##1}}}
\expandafter\def\csname PY@tok@nt\endcsname{\let\PY@bf=\textbf\def\PY@tc##1{\textcolor[rgb]{0.00,0.50,0.00}{##1}}}
\expandafter\def\csname PY@tok@nv\endcsname{\def\PY@tc##1{\textcolor[rgb]{0.10,0.09,0.49}{##1}}}
\expandafter\def\csname PY@tok@s1\endcsname{\def\PY@tc##1{\textcolor[rgb]{0.73,0.13,0.13}{##1}}}
\expandafter\def\csname PY@tok@dl\endcsname{\def\PY@tc##1{\textcolor[rgb]{0.73,0.13,0.13}{##1}}}
\expandafter\def\csname PY@tok@ch\endcsname{\let\PY@it=\textit\def\PY@tc##1{\textcolor[rgb]{0.25,0.50,0.50}{##1}}}
\expandafter\def\csname PY@tok@m\endcsname{\def\PY@tc##1{\textcolor[rgb]{0.40,0.40,0.40}{##1}}}
\expandafter\def\csname PY@tok@gp\endcsname{\let\PY@bf=\textbf\def\PY@tc##1{\textcolor[rgb]{0.00,0.00,0.50}{##1}}}
\expandafter\def\csname PY@tok@sh\endcsname{\def\PY@tc##1{\textcolor[rgb]{0.73,0.13,0.13}{##1}}}
\expandafter\def\csname PY@tok@ow\endcsname{\let\PY@bf=\textbf\def\PY@tc##1{\textcolor[rgb]{0.67,0.13,1.00}{##1}}}
\expandafter\def\csname PY@tok@sx\endcsname{\def\PY@tc##1{\textcolor[rgb]{0.00,0.50,0.00}{##1}}}
\expandafter\def\csname PY@tok@bp\endcsname{\def\PY@tc##1{\textcolor[rgb]{0.00,0.50,0.00}{##1}}}
\expandafter\def\csname PY@tok@c1\endcsname{\let\PY@it=\textit\def\PY@tc##1{\textcolor[rgb]{0.25,0.50,0.50}{##1}}}
\expandafter\def\csname PY@tok@fm\endcsname{\def\PY@tc##1{\textcolor[rgb]{0.00,0.00,1.00}{##1}}}
\expandafter\def\csname PY@tok@o\endcsname{\def\PY@tc##1{\textcolor[rgb]{0.40,0.40,0.40}{##1}}}
\expandafter\def\csname PY@tok@kc\endcsname{\let\PY@bf=\textbf\def\PY@tc##1{\textcolor[rgb]{0.00,0.50,0.00}{##1}}}
\expandafter\def\csname PY@tok@c\endcsname{\let\PY@it=\textit\def\PY@tc##1{\textcolor[rgb]{0.25,0.50,0.50}{##1}}}
\expandafter\def\csname PY@tok@mf\endcsname{\def\PY@tc##1{\textcolor[rgb]{0.40,0.40,0.40}{##1}}}
\expandafter\def\csname PY@tok@err\endcsname{\def\PY@bc##1{\setlength{\fboxsep}{0pt}\fcolorbox[rgb]{1.00,0.00,0.00}{1,1,1}{\strut ##1}}}
\expandafter\def\csname PY@tok@mb\endcsname{\def\PY@tc##1{\textcolor[rgb]{0.40,0.40,0.40}{##1}}}
\expandafter\def\csname PY@tok@ss\endcsname{\def\PY@tc##1{\textcolor[rgb]{0.10,0.09,0.49}{##1}}}
\expandafter\def\csname PY@tok@sr\endcsname{\def\PY@tc##1{\textcolor[rgb]{0.73,0.40,0.53}{##1}}}
\expandafter\def\csname PY@tok@mo\endcsname{\def\PY@tc##1{\textcolor[rgb]{0.40,0.40,0.40}{##1}}}
\expandafter\def\csname PY@tok@kd\endcsname{\let\PY@bf=\textbf\def\PY@tc##1{\textcolor[rgb]{0.00,0.50,0.00}{##1}}}
\expandafter\def\csname PY@tok@mi\endcsname{\def\PY@tc##1{\textcolor[rgb]{0.40,0.40,0.40}{##1}}}
\expandafter\def\csname PY@tok@kn\endcsname{\let\PY@bf=\textbf\def\PY@tc##1{\textcolor[rgb]{0.00,0.50,0.00}{##1}}}
\expandafter\def\csname PY@tok@cpf\endcsname{\let\PY@it=\textit\def\PY@tc##1{\textcolor[rgb]{0.25,0.50,0.50}{##1}}}
\expandafter\def\csname PY@tok@kr\endcsname{\let\PY@bf=\textbf\def\PY@tc##1{\textcolor[rgb]{0.00,0.50,0.00}{##1}}}
\expandafter\def\csname PY@tok@s\endcsname{\def\PY@tc##1{\textcolor[rgb]{0.73,0.13,0.13}{##1}}}
\expandafter\def\csname PY@tok@kp\endcsname{\def\PY@tc##1{\textcolor[rgb]{0.00,0.50,0.00}{##1}}}
\expandafter\def\csname PY@tok@w\endcsname{\def\PY@tc##1{\textcolor[rgb]{0.73,0.73,0.73}{##1}}}
\expandafter\def\csname PY@tok@kt\endcsname{\def\PY@tc##1{\textcolor[rgb]{0.69,0.00,0.25}{##1}}}
\expandafter\def\csname PY@tok@sc\endcsname{\def\PY@tc##1{\textcolor[rgb]{0.73,0.13,0.13}{##1}}}
\expandafter\def\csname PY@tok@sb\endcsname{\def\PY@tc##1{\textcolor[rgb]{0.73,0.13,0.13}{##1}}}
\expandafter\def\csname PY@tok@sa\endcsname{\def\PY@tc##1{\textcolor[rgb]{0.73,0.13,0.13}{##1}}}
\expandafter\def\csname PY@tok@k\endcsname{\let\PY@bf=\textbf\def\PY@tc##1{\textcolor[rgb]{0.00,0.50,0.00}{##1}}}
\expandafter\def\csname PY@tok@se\endcsname{\let\PY@bf=\textbf\def\PY@tc##1{\textcolor[rgb]{0.73,0.40,0.13}{##1}}}
\expandafter\def\csname PY@tok@sd\endcsname{\let\PY@it=\textit\def\PY@tc##1{\textcolor[rgb]{0.73,0.13,0.13}{##1}}}

\makeatother

    \definecolor{incolor}{rgb}{0.0, 0.0, 0.5}
    \definecolor{outcolor}{rgb}{0.545, 0.0, 0.0}

    \hypersetup{
      breaklinks=true,  
      colorlinks=true,
      urlcolor=urlcolor,
      linkcolor=linkcolor,
      citecolor=citecolor,
      }
    
    \usepackage{pstricks}

\usepackage{amsmath,amstext,amssymb,amsfonts}
\usepackage{graphicx}
\usepackage{xcolor}
\usepackage{tabularx}
\usepackage{verbatim}
\usepackage{fullpage}
\usepackage[T1]{fontenc}
\usepackage{bbm}

\usepackage{amsthm}
\usepackage{ifdraft}
\usepackage{algorithm2e}
\usepackage{algorithmic}
\usepackage{appendix}
\usepackage{multicol}

\usepackage{nicefrac}

\newcommand{\flatfrac}[2]{#1/#2}
\newcommand{\ffrac}{\flatfrac}

\usepackage{microtype}

\newtheorem{theorem}{Theorem}
\newtheorem*{theorem*}{Theorem}

\newtheorem*{claim*}{Claim}

\newtheorem{proposition}[theorem]{Proposition}
\newtheorem*{proposition*}{Proposition}
\newtheorem{lemma}[theorem]{Lemma}
\newtheorem*{lemma*}{Lemma}
\newtheorem{corollary}[theorem]{Corollary}

\newtheorem*{conjecture*}{Conjecture}
\newtheorem{observation}[theorem]{Observation}
\newtheorem{fact}[theorem]{Fact}
\newtheorem*{fact*}{Fact}

\newtheorem*{hypothesis*}{Hypothesis}

\theoremstyle{definition}
\newtheorem{definition}[theorem]{Definition}

\newtheorem{SDP}[theorem]{SDP}

\newtheorem{QP}[theorem]{QP}
\newtheorem{LP}[theorem]{LP}

\usepackage{prettyref}
\newcommand{\savehyperref}[2]{\texorpdfstring{\hyperref[#1]{#2}}{#2}}

\newrefformat{parta}{\savehyperref{#1}{a}}
\newrefformat{partb}{\savehyperref{#1}{b}}
\newrefformat{eq}{\savehyperref{#1}{\textup{(\ref*{#1})}}}
\newrefformat{lem}{\savehyperref{#1}{Lemma~\ref*{#1}}}
\newrefformat{def}{\savehyperref{#1}{Definition~\ref*{#1}}}
\newrefformat{thm}{\savehyperref{#1}{Theorem~\ref*{#1}}}
\newrefformat{cor}{\savehyperref{#1}{Corollary~\ref*{#1}}}
\newrefformat{cha}{\savehyperref{#1}{Chapter~\ref*{#1}}}
\newrefformat{sec}{\savehyperref{#1}{Section~\ref*{#1}}}
\newrefformat{app}{\savehyperref{#1}{Appendix~\ref*{#1}}}
\newrefformat{tab}{\savehyperref{#1}{Table~\ref*{#1}}}
\newrefformat{fig}{\savehyperref{#1}{Figure~\ref*{#1}}}
\newrefformat{hyp}{\savehyperref{#1}{Hypothesis~\ref*{#1}}}
\newrefformat{alg}{\savehyperref{#1}{Algorithm~\ref*{#1}}}
\newrefformat{sdp}{\savehyperref{#1}{SDP~\ref*{#1}}}
\newrefformat{qp}{\savehyperref{#1}{QP~\ref*{#1}}}
\newrefformat{vp}{\savehyperref{#1}{VP~\ref*{#1}}}
\newrefformat{lp}{\savehyperref{#1}{LP~\ref*{#1}}}
\newrefformat{rem}{\savehyperref{#1}{Remark~\ref*{#1}}}
\newrefformat{item}{\savehyperref{#1}{Item~\ref*{#1}}}
\newrefformat{step}{\savehyperref{#1}{step~\ref*{#1}}}
\newrefformat{conj}{\savehyperref{#1}{Conjecture~\ref*{#1}}}
\newrefformat{fact}{\savehyperref{#1}{Fact~\ref*{#1}}}
\newrefformat{prop}{\savehyperref{#1}{Proposition~\ref*{#1}}}
\newrefformat{claim}{\savehyperref{#1}{Claim~\ref*{#1}}}
\newrefformat{relax}{\savehyperref{#1}{Relaxation~\ref*{#1}}}
\newrefformat{red}{\savehyperref{#1}{Reduction~\ref*{#1}}}
\newrefformat{part}{\savehyperref{#1}{Part~\ref*{#1}}}
\newrefformat{prob}{\savehyperref{#1}{Problem~\ref*{#1}}}
\newrefformat{ass}{\savehyperref{#1}{Assumption~\ref*{#1}}}
\newrefformat{cons}{\savehyperref{#1}{Construction~\ref*{#1}}}

\newcommand{\Sref}[1]{\hyperref[#1]{\S\ref*{#1}}}

\renewcommand{\leq}{\leqslant}

\renewcommand{\geq}{\geqslant}

\usepackage{bm}

\usepackage{xspace}

\usepackage{boxedminipage}

\newenvironment{mybox}
{\center \noindent\begin{boxedminipage}{1.0\linewidth}}
{\end{boxedminipage}
\noindent
}

\newcommand{\mper}{\,.}

\newcommand{\paren}[1]{\left(#1 \right )}

\newcommand{\Brac}[1]{\left[#1\right]}

\newcommand{\set}[1]{\left\{#1\right\}}

\newcommand{\abs}[1]{\left\lvert#1\right\rvert}

\newcommand{\norm}[1]{\left\lVert#1\right\rVert}

\newcommand{\defeq}{\stackrel{\textup{def}}{=}}

\newcommand{\inprod}[1]{\left\langle #1\right\rangle}

\newcommand{\R}{\mathbb R}

\newcommand{\OPT}{{\sf OPT}}

\newcommand{\subjectto}{\text{subject to}}

\newcommand{\Esymb}{\mathbb{E}}
\newcommand{\Psymb}{\mathbb{P}}

\DeclareMathOperator*{\E}{\Esymb}

\DeclareMathOperator*{\ProbOp}{\Psymb}

\renewcommand{\Pr}[1]{\ProbOp\Brac{#1}}

\newcommand{\e}{\epsilon}

\newcommand{\one}{\mathbbm{1}}

\definecolor{DSgray}{cmyk}{0,0,0,0.7}

\let\e\varepsilon

\newcommand{\cE}{\mathcal E}
\newcommand{\cF}{\mathcal F}

\newcommand{\bbR}{\mathbb R}

\newcommand{\Erdos}{Erd\H{o}s\xspace}
\newcommand{\Renyi}{R\'enyi\xspace}

\newcommand{\bigO}{\mathcal{O}}

\newcommand{\poly}{{\sf poly}}

\newcommand{\rank}{{\sf rank}}

\newcommand\numberthis{\addtocounter{equation}{1}\tag{\theequation}} 

\newcommand{\restrict}[2]{\left.{#1}\right|_{{#2}}}

\newcommand{\SDPOPT}{{\sf SDPOPT}}

\author{
	Akash Kumar\footnote{{\'Ecole Polytechnique F\'ed\'erale de Lausanne}, Lausanne, Switzerland.}\\ \href{mailto:akash.kumar@epfl.ch}{akash.kumar@epfl.ch}
	\and 
	Anand Louis\footnote{Indian Institute of Science, Bangalore, India.}\\ \href{mailto:anandl@iisc.ac.in}{anandl@iisc.ac.in}
	\and 
	Rameesh Paul\footnotemark[2]\\ \href{mailto:rameeshpaul@iisc.ac.in}{rameeshpaul@iisc.ac.in}
}

\title{Exact recovery algorithm for Planted Bipartite Graph in Semi-random Graphs}

\begin{document}     
\maketitle
\begin{abstract}
The problem of finding the largest induced balanced bipartite subgraph in a given graph is NP-hard. This problem is closely related to the problem of finding the smallest Odd Cycle Transversal. 

In this work, we consider the following model of instances: starting with a set of vertices $V$, a set $S \subseteq V$ of $k$ vertices is chosen and an arbitrary $d$-regular bipartite graph is added on it; edges between pairs of vertices in $S\times \paren{V\setminus S}$ and $\paren{V\setminus S} \times \paren{V\setminus S}$ are added with probability $p$. Since for $d=0$, the problem reduces to recovering a planted independent set, we don't expect efficient algorithms for $k=o\paren{\sqrt{n}}$. This problem is a generalization of the planted balanced biclique problem where the bipartite graph induced on $S$ is a complete bipartite graph;
\cite{Lev18} gave an algorithm for recovering $S$ in this problem when $k=\Omega\paren{\sqrt{n}}$.

Our main result is an efficient algorithm that recovers (w.h.p.) the planted bipartite graph when $k=\Omega_p\paren{\sqrt{n \log n}}$ for a large range of parameters. Our results also hold for a natural semi-random model of instances, which involve the presence of a monotone adversary. Our proof shows that a natural SDP relaxation for the problem is integral by constructing an appropriate solution to it's dual formulation. Our main technical contribution is a new approach for constructing the dual solution where we calibrate the eigenvectors of the adjacency matrix to be the eigenvectors of the dual matrix. We believe that this approach may have applications to other recovery problems in semi-random models as well.

When $k=\Omega\paren{\sqrt{n}}$, we give an algorithm for recovering $S$ whose running time is exponential in the number of small eigenvalues in graph induced on $S$; this algorithm is based on subspace enumeration techniques due to the works of \cite{KT07,ABS10,Kol11}.
\end{abstract}
\newpage 
\tableofcontents
\newpage
\section{Introduction}
Given a graph $G=\paren{V,E}$, the problem of finding the largest induced bipartite subgraph of $G$ is well known to be NP-hard \cite{Yan78}. The problem is equivalent to the Odd Cycle Transversal problem. The problem is also related to the balanced biclique problem, where the task is that of finding the largest induced balanced complete bipartite subgraph.  
This problem has a lot of practical application in computational biology \cite{CC00}, bioinformatics  \cite{Zha08} and VLSI design \cite{AM99}. 

For the worst-case instance of the problem, the work \cite{ACMM05} gives an algorithm that computes a set with at least $\paren{1-\bigO\paren{\e\sqrt{\log n}}}$ fraction of vertices which induces a bipartite graph, when it is promised that the graph contains an induced bipartite graph having $\paren{1-\varepsilon}n$ fraction of the vertices.
The work \cite{GL21} gives an efficient randomized algorithm that computes an induced bipartite subgraph having $\paren{1-\bigO\paren{\sqrt{\varepsilon \log d}}}$ fraction of the vertices where $d$ is the bound on the maximum degree of the graph. They also give a matching (up to constant factors) Unique Games hardness for certain regimes of parameters. We refer to \prettyref{sec:related_work} for more details about these related problems.

In an effort to better understand the complexity of various computationally intractable problems, a lot of work has been focused on the special cases of the problem, and towards studying the problem in various \textit{random} and \textit{semi-random} models. Here, one starts with solving the problem for random instances (for graph problems this is often $G_{n,p}$ \Erdos-\Renyi graphs\footnote{For each pair of vertices, an edge is added independently with probability $p$.}). The analysis in random instances is often much simpler, and one can give algorithms with \say{good} approximation guarantees. The next goal in this direction is to plant a solution that is \say{clearly optimal} in an ambient random graph and then attempt to recover this planted solution. We, therefore, build towards the worst-case instances of the problem by progressively weakening our assumptions. We refer to the book \cite{Rou21} for a more detailed discussion of these models in the context of other problems like planted clique, planted bisection, $k$-coloring, Stochastic Block Models, and Matrix completion problems.

We start our discussion with the problem of computing a maximum clique/independent set, since it has been extensively studied in such planted models. In the planted clique/independent set problem we plant a clique/independent set of size $k$ in an otherwise random $G_{n,p}$ graph. The work \cite{AKS98} presents an algorithm, which, given a graph $G \sim \mathcal{G}(n, 1/2)$ with a planted clique/independent set of size $k$, recovers the planted clique when $k > c_1\sqrt{n}$ (where $c_1$ is a constant). We will refer to the planted independent set/clique problem at various points throughout the introduction

Such \textit{random planted models} have been studied in context of other problems as well such as the planted $3$-coloring problem \cite{BS95,AK97}, planted dense subgraph problem \cite{HWX16a,HWX16b,HWX16c}, planted bisection and planted Stochastic Block models \cite{BSFM87,DF89,JS98,CI01,CK01,ABH16}, to state a few. We define a similar \textit{random planted model} to study our problem, as stated below.

\begin{definition} [Random planted model]
\label{def:planted_model}
Given $n,k,d,p$, our planted bipartite graph is constructed as follows,
\begin{enumerate}
    \item Let $V$ be a set of $n$ vertices. Fix an arbitrary subset $S \subset V$ such that $\abs{S}=k$.
    \item Add edges arbitrarily inside $S$ such that the resulting graph is a connected $d$-regular bipartite graph. Let $S_1,S_2$ denote the bipartite components.
    \item For each pair of vertices in $S \times \paren{V\setminus S}$, add an edge independently with probability $p$.
    \item For each pair of vertices in $\paren{V \setminus S} \times \paren{V\setminus S}$, add an edge independently with probability $p$.
\end{enumerate}
\end{definition}

For planted cliques, a lot of work has been done in the special case of $p=1/2$. However, people have studied other problems such as the planted bisection problems \cite{FK01}, and exact recovery problems in Stochastic Block Models \cite{ABH16} in the harder $p=o(1)$ regimes. Therefore, we also aim to solve our problem in $p=o(1)$ regimes.

We note that this problem is a generalization of the planted independent set and the planted balanced biclique problem. For $d=0$, it reduces to recovering a planted independent set and hence we do not expect efficient algorithms for $k=o\paren{\sqrt{n}}$ \cite{FGR+13,BHK+16}. For $k =\Omega\paren{\sqrt{n}}$, both these special cases i.e the planted independent set problem \cite{AKS98,FK00}, and the planted balanced biclique problem \cite{Lev18} admit a polynomial-time recovery algorithm. So it is natural to consider  $k=\Omega\paren{\sqrt{n}}$ as a benchmark for recovery and look for algorithms in this regime. 
The other consideration for interesting regimes to study the problem comes by viewing this problem as a special case of the densest $k$-subgraph (DkS) problem. When $d \gg pk$, the problem can be viewed as the densest $k$-subgraph (DkS) and for $d \ll pk$, the problem can be viewed as sparsest $k$-subgraph problem (studying the complement of this graph would be an instance of DkS problem). However, this general DkS problem is information-theoretically unsolvable for $d=pk$ \cite{CX16}. Formally, this follows from Theorem 2.1 in the work \cite{CX16}, by setting $d=qk=pk$ and setting $r=1$ where $q$ is the edge probability within the vertices of planted subgraph and a $p$ is the edge probability when at least one of the vertex does not belong to the planted subgraph and $r$ is the number of clusters. Therefore we focus our attention to the case when $d \approx pk$ (also including $d=pk$). In our problem, we can hope to use the specifics of the bipartite structure in hand and recover the planted set exactly. 
\subsection{Our models and results} \label{sec:models_results}
\begin{figure}[htbp]
\centerline{\includegraphics[scale=.4]{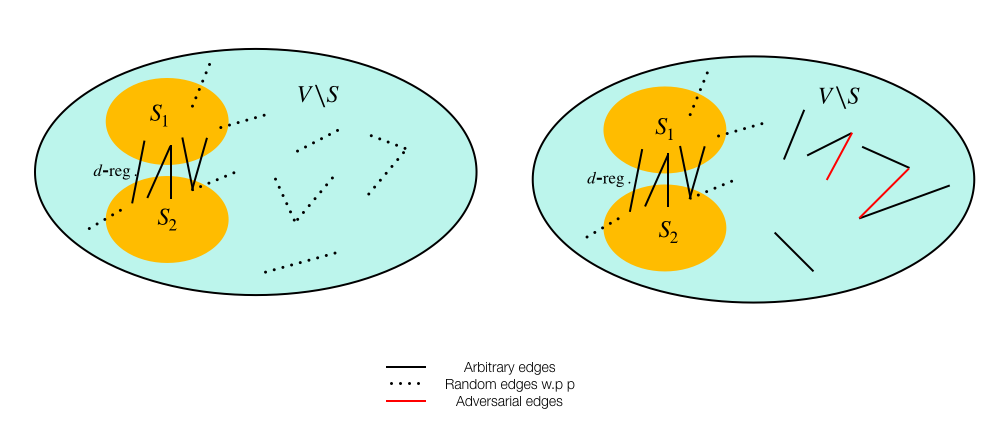}}
\caption{Random Planted model \prettyref{def:planted_model} (left) and semi-random model \prettyref{def:semi-random_model} (right).}
\label{fig:a}
\end{figure}

We start by introducing our semi-random model which attempts to robustify the random planted model from \prettyref{def:planted_model}.

\begin{definition} [Semi-random model]
\label{def:semi-random_model}
Fix $n,k,d,p$, we now describe how a graph $G$ from our semi-random model is generated as,
\begin{enumerate}
     \item Let $V$ be a set of $n$ vertices. Fix an arbitrary subset $S \subset V$ such that $\abs{S}=k$.
    \item Add edges arbitrarily inside $S$ such that the resulting graph is a connected $d$-regular bipartite graph. Let $S_1,S_2$ denote the bipartite components.
    \item For each pair of vertices in $S \times \paren{V\setminus S}$, add an edge independently with probability $p$.
     \item Arbitrarily add edges in $\paren{V \setminus S} \times \paren{V \setminus S}$ such that
     smallest eigenvalue of the matrix $\paren{A_{\paren{V \setminus S} \times \paren{V \setminus S}}-p\one_{V \setminus S}\one_{V \setminus S}^T}$ is greater than
     $-\paren{\ffrac{(1/2-\alpha)}{(1/2+\alpha)}}d$ where $\alpha$ is a small\footnote{Note that the smaller the value of $\alpha$, the weaker is this assumption.} positive constant (throughout this paper we assume $\alpha \leq \ffrac{1}{6}$).
    \label{step:rest_of_graph}
    \item  \label{step:adversary} Allow a {monotone adversary} to add edges in $\paren{V\setminus S} \times \paren{V \setminus S}$ arbitrarily.
\end{enumerate}
\end{definition}

\begin{observation}\prettyref{def:semi-random_model} also captures \prettyref{def:planted_model}; since in the case when $V\setminus$S is chosen to be a $G_{\paren{n-k},p}$ random graph, $\paren{A_{\paren{V \setminus S} \times \paren{V \times S}} -p \one_{V \setminus S}\one_{V \setminus S}^T} = A_{\paren{V \setminus S} \times \paren{V \setminus S}} - \E\Brac{A_{\paren{V \setminus S} \times \paren{V \setminus S}}}$, and therefore the smallest eigenvalue of $\paren{A_{\paren{V \setminus S} \times \paren{V \times S}} -p \one_{V \setminus S}\one_{V \setminus S}^T}$ is greater than $-2\sqrt{n}$ (as follows from the work \cite{Vu07}).
\end{observation}

Models stronger than \textit{random planted models} have also been considered in the literature for planted problems. The work \cite{FK00} studies the planted clique problem in what they call the \say{sandwich model}. The model is constructed as per the random planted model in  \prettyref{def:planted_model}, but an adversary is allowed to act on the top of that in a fashion similar to \prettyref{step:adversary} of \prettyref{def:semi-random_model}.

The work \cite{FK01} introduced a strong adversarial \textit{semi-random} model (referred to as the \emph{Fiege and Kilian} model). They gave recovery algorithms for the planted clique ($k=\Omega(n)$ regimes) and for the planted bisection and planted $k$-coloring in this model. The work \cite{MMT20} further shows that one can recover the planted clique for $k = \Omega_p\paren{n^{2/3}}$ \footnote{$\Omega_{p}$ hides $\poly\paren{\ffrac{1}{p}}$ factors.} in \cite{FK01} model. 

In the Feige-Kilian model, \prettyref{step:rest_of_graph} allows for any arbitrary graph in $\paren{V \setminus S} \times \paren{V \setminus S}$. However, with no further assumptions on graph induced on $V \setminus S$, even for the special case of planted independent set problem ($d=0$), the best known algorithm \cite{MMT20} works only for $k=\Omega_p\paren{n^{\ffrac{2}{3}}}$. However, since our benchmark is $k=\Omega\paren{\sqrt{n}}$, we look at a model with stronger assumptions than the Feige-Kilian model. In order to uniquely identify the planted graph, we need to assume that $V \setminus S$ is far from having any induced bipartite subgraphs of degree at least $d$.
Our condition in \prettyref{step:rest_of_graph} implies that this indeed holds. This is because if the smallest eigenvalue is greater than $-d/2+2\sqrt{n}$, the graph is indeed far from having an induced bipartite subgraph of smallest degree $d$. Since otherwise, a vector having entries $1$ for one side of the bipartition and $-1$ on the other side and $0$ elsewhere achieves a Rayleigh Quotient of value $-d$ (and hence the smallest eigenvalue is at most $-d$). 

We now present our main result which holds for both the random planted model (\prettyref{def:planted_model}) and semi-random model (\prettyref{def:semi-random_model}).

\begin{theorem}[Informal version of \prettyref{thm:arbitrary_formal} ]
\label{thm:arbitrary_informal}
For $n,k,d,p$ satisfying $k = \Omega_p\paren{\sqrt{n \log n}}$ and $p=\Omega\paren{\log k/k}^{1/6}$ and $d \geq 2pk/3$, there exists a deterministic algorithm that takes as input an instance generated by \prettyref{def:semi-random_model}, and recovers the arbitrary planted set $S$ exactly, in polynomial time and with high probability (over the randomness of the input).
\end{theorem}

Achieving exact recovery for $k=\Omega_p\paren{\sqrt{n}}$ is still an open problem. To the best of our knowledge, nothing is known about this problem in full generality.
For the planted clique problem, recovery for $k=\Omega\paren{\sqrt{n \log n}}$ is trivial \cite{Kuc95}. However, such techniques don't work for our problem when $d=pk$. We prove \prettyref{thm:arbitrary_informal} by showing that an SDP relaxation for the problem is integral, by constructing an optimal dual solution. We give an outline of the proof in \prettyref{sec:proof_overview} and a detailed proof in \prettyref{sec:sdp_section}. 

Our proofs use the spectral properties of bipartite graphs and random graphs to show the existence of an optimal dual solution having large rank.
Our main technical contribution is a new approach for constructing a dual solution where we \emph{calibrate the eigenvectors} of the adjacency matrix to be the eigenvectors of the dual matrix. We believe that this approach may have applications to other recovery problems in semi-random models as well.

\begin{theorem}[Informal version of \prettyref{thm:subspace_formal}]
\label{thm:subspace_informal}
For $n,k,d,p$, satisfying $k=\Omega_p\paren{\sqrt{n}}$, there exists a deterministic algorithm that takes as input an instance generated as per \prettyref{def:planted_model}, and recovers the arbitrary planted set $S$ exactly with high probability (over the randomness of the input) in time exponential in the number of small eigenvalues of the adjacency matrix (eigenvalues smaller than $-d/2+2\sqrt{n}$) of the graph induced on $S$.
\end{theorem}

\begin{observation}\label{remark:poly_alg_cases}
For and many special classes of instances such as, (i) 
when the probability $p=\Omega\paren{1}$, (ii) when the planted graph is a complete bipartite graph like in the balanced biclique problem (iii) when the planted bipartite graph is a $d$-regular random graph or (iv) more generally when the planted graph is a $d$-regular expander graph; the number of these small eigenvalues is a constant in the regimes of $d=\Omega(pk)$ and \prettyref{thm:subspace_informal} allows efficient recovery (running time of the algorithm is polynomial in $n$).
\end{observation}

\subsection{Related Work} \label{sec:related_work}
\paragraph*{Odd Cycle Transversal problem}
The odd cycle transversal problem asks to find the smallest set of vertices in the graph such that the set has an intersection with every odd cycle of the graph. Removing these vertices will result in a bipartite graph, and hence this problem is equivalent to finding the largest induced bipartite graph. Owing to the hereditary nature of the bipartiteness property, the problem is NP-hard, as follows from the work of Yannakakis \cite{Yan78}. The work \cite{Yan78} shows that for a broad class of problems that have a structure that is hereditary on induced subgraphs, finding such a structure is NP-Complete.
The optimal long code test by Khot and Bansal \cite{BK09} rules out any constant factor approximation for this problem. On the algorithmic front, casting the problem as a $2$-\textsf{CNF} deletion problem, \cite{AKRR90} gives a reduction to the min-multicut problem. This reduction gives us an $\bigO\paren{\log n}$ approximation due to the work \cite{GVY98}, which was further improved to $\bigO\paren{\sqrt{\log n}}$ in the work \cite{ACMM05}.
The work \cite{GL21} gives an efficient randomized algorithm that removes only $\bigO\paren{n\sqrt{\OPT \log d}}$ vertices where $d$ is the bound on the maximum degree of the graph and $\OPT$ denotes the fraction of vertices in the optimal set.  They also give a matching (up to constant factors)
Unique Games hardness for certain regimes of parameters.

The problem is equivalent to finding the largest 2-colorable subgraph of a given graph and is known as the partial 2-coloring problem. The work \cite{GLR18} studies the problem in the Feige-Kilian semi-random model \cite{FK01}, where a 2-colorable graph of size $\paren{1-\varepsilon}n$ is planted. They give an algorithm that outputs a set $\mathcal{S}'$ such that $\abs{\mathcal{S'}} \geq \paren{1-\ffrac{\varepsilon c}{p^2}}n$ for $p=\Omega\paren{\sqrt{\ffrac{\log n}{n}}}$ and $\varepsilon \leq p^2$ where $c$ is a positive constant. 
Their algorithm is a partial recovery algorithm and works for the regimes when $\varepsilon$ is small.
Our results in \prettyref{thm:arbitrary_informal} hold when $1-\varepsilon$ is small and give complete recovery for a large range of $p$.
However, since our model in \prettyref{def:semi-random_model} makes stronger assumptions than the \cite{FK01} model, we don't make any comparisons.

\paragraph*{Balanced Biclique problem}
In the balanced complete bipartite subgraph problem (also called the balanced biclique problem), we are given a graph on $n$ vertices and a parameter $k$, and the problem then asks whether there is a complete bipartite subgraph that is balanced with $k$ vertices in each of the bipartite components. The problem was studied when the underlying graph is a bipartite graph, and shown to be NP-complete by a reduction from the CLIQUE problem in the works \cite{GJ79,Joh87}. They additionally note that the balanced constraint is what makes the problem hard. If we remove the balanced constraint, the problem can be reduced to finding a maximum independent set in a bipartite graph. The latter problem admits a polynomial-time solution using the matching algorithm. The work \cite{FK04} shows that this problem of finding a maximum balanced biclique is hard to approximate within a factor of $2^{\paren{\log n}^{\delta}}$ for some $\delta >0$, under the assumption that $\mathsf{3SAT} \notin \textsf{DTIME}\paren{2^{n^{3/4+\varepsilon}}}$ for some $\varepsilon >0$. Recently, the work \cite{Man17} showed that one cannot find a better approximation than $n^{1-\varepsilon}$, assuming the \textit{Small Set Expansion Hypothesis} and that $\textsf{NP} \nsubseteq \textsf{BPP}$ for every constant $\varepsilon >0$.

A related problem is the maximum edge biclique problem, where we are asked to find whether $G$ contains a biclique with at least $k$ edges. This problem was also shown to be NP-hard in the work \cite{Pee03}. 

Given these intractability results for general graphs, there has been some success in special classes of graphs. In graphs with constant arboricity, the work \cite{Epp94} gives a linear time algorithm that lists all maximal complete bipartite subgraphs. In a degree bounded graph, the work \cite{TSS02} gives a combinatorial algorithm for the balanced biclique problem that runs in time $\bigO\paren{n2^d}$.  Another systematic approach, however, is to consider planted and semi-random models for the problem. In the work \cite{Lev18}, they study the planted version of the problem, which, they call \say{hidden biclique problem}. Their model is similar to our model in \prettyref{def:planted_model}; however, we consider an arbitrary $d$-regular bipartite graph instead of a complete bipartite graph. They give a linear-time combinatorial algorithm that finds the planted hidden biclique with high probability (over the randomness of the input instance) for $k=\Omega\paren{\sqrt{n}}$. Their algorithm builds on the \say{Low Degree Removal} algorithm, due to Feige and Ron \cite{FR10} which finds a planted clique in linear time.

\paragraph*{Graph problems in Semi-random and Pseudorandom models}
A wide variety of random graph models and their relaxations have been a rich source of algorithmic problems on graphs. Alon and Kahale \cite{AK97} sharpened the results of Blum and Spencer \cite{BS95} and gave algorithms that recover a planted $3$-coloring in a natural family of random $3$-colorable instances. \cite{KLT17} extended this result and showed how to recover a $3$-coloring when the input graph is pseudorandom (has some mild expansion properties) and is known to admit a random like $3$-coloring. A unified spectral approach by McSherry \cite{McS01} gives a single shot recovery algorithm for many problems in these random planted models. One can use the \cite{McS01} framework to recover a planted random bipartite graph; however, it is not known if it will work if $S$ is an arbitrary bipartite graph.

On the other side, we have semi-random models. Notably, the Feige-Kilian model \cite{FK01} is one of the strongest semi-random models. In \cite{FK01}, they also give recovery algorithms for planted clique, planted $k$-colorable, and planted bisection problem in this model. In \cite{MMT20}, they give a recovery algorithm for the independent set problem for large regimes of parameters. The work \cite{KLP21} generalizes these results to $r$-uniform hypergraphs in this model. There are other works \cite{MMV12,MMV14,LV18,LV19} that study graph partitioning in semi-random models.

A host of work has been done in various random and semi-random models for the more general densest $k$-subgraph problem. The works by Hajek, Wu, and Xu \cite{HWX16a,HWX16b,HWX16c} study the problem when the planted dense subgraph is random and gives algorithms for exact recovery using SDP relaxations for some range of parameters. They complement these results by providing information-theoretic limits for regimes where recovery is impossible. The work by \cite{BCC+10} studies this problem when the planted graph is arbitrary. They analyze an SDP-based method to distinguish the dense graphs from the family of $G_{n,p}$ graphs when $k \geq \sqrt{n}$. The work \cite{KL20} studies the problem of densest $k$-subgraph in some semi-random model and gives a partial recovery algorithm for some regimes of $d,k,n,p$.

SDP has been the tool of choice for exact recovery in semi-random models. Starting from the fundamental works of exact recovery for the planted clique problem \cite{FK00}, for the planted bisection problem \cite{FK01}, for Stochastic Block Models \cite{ABH16} etc., (and many other works as have been mentioned above), are based on SDP relaxations. A natural way to analyze these SDP relaxations is by constructing an optimal dual solution to prove integrality of the primal relaxation. This idea has been explored in the works of \cite{FK01,CO07,BCC+10,ABBS14,ABH16,LV18}, to state a few. We note that the task of constructing an optimal dual solution is problem-specific, and there is no generic way of doing this.

\subsection{Preliminaries}
We start with some essential notation to understand the proof overview and review some well-known facts about random perturbation matrices. Then, we write our SDP relaxation to the problem and the accompanying dual SDP. We follow this up with a discussion on some well known tools from spectral graph theory such as the \emph{threshold rank} and \emph{spectral embedding}. We will build on these ideas in our Proof Overview \prettyref{sec:proof_overview} to show that the primal SDP is an optimal one and the primal matrix is a rank-one matrix.

\subsubsection{Notation}
We let $\Brac{M}_{n \times n}$ denote a matrix $M$ of size $n \times n$. For some set of indices $R_1,R_2 \subseteq [n]$, $M_{R_1 \times R_2}$  denotes a matrix of size $n\times n$ constructed out of matrix $M$ of size $n \times n$ by copying the entries for $(i,j) \in R_1 \times R_2$ and setting rest of the entries to be $0$.
We let $\left.M\right|_{R_1 \times R_2}$  denote the matrix of size $\abs{R_1} \times \abs{R_2}$ constructed from a matrix $M$ of size $n \times n$ by taking rows corresponding to $R_1$ and columns corresponding to $R_2$.
The eigenvalues of a matrix $M$ are sorted as $\lambda_1(M) \leq \lambda_2(M) \leq \hdots \leq \lambda_n(M)$. We will drop the matrix $M$ wherever it is clear from the context. The eigenvectors are also sorted by their corresponding eigenvalues.

\subsubsection{Spectral bounds on Perturbation matrices}
We let $A$ denote the adjacency matrix of the graph obtained using \prettyref{def:planted_model}.
We can express the matrix $A$ as sum of \say{simpler} matrices,
\begin{align} \label{eq:matrix_express}
    A = A_{S \times S} + A_{V \setminus S \times V \setminus S} + p\paren{\one\one^T-\one_S\one_{S}^T-\one_{V\setminus S}\one_{V \setminus S}^T} + R && \paren{ R_{ij} \defeq A_{ij}-\E[A_{ij}]}
\end{align}
 where $A_{S\times S}$ represents the matrix corresponding to the planted bipartite graph, the term  $p\paren{\one\one^T-\one_S\one_{S}^T - \one_{V\setminus S}\one_{V \setminus S}^T}$ is the expected adjacency matrix for the random graph and $R$ as defined above is the perturbation matrix corresponding to the random part of the graph. 

\begin{proposition}\label{claim:random_matrix_norm}
For the perturbation matrix $R$ as defined in equation \prettyref{eq:matrix_express} we have that $\norm{R} \leq 2\sqrt{n}$ almost surely.
\end{proposition}

\begin{proof}
$R$ is a symmetric random matrix and the entries $R_{ij}$ can be treated as random variables, bounded between $-1$ and $1$, with expectation $0$ and variance $p\paren{1-p} \leq \ffrac{1}{4}$. Also the entries $R_{ij}$ are independent and hence, by Theorem 1.1 in the work \cite{Vu07}, we have $\norm{R} \leq 2\sqrt{n}$ almost surely.
\end{proof}

\subsubsection{SDP Relaxation}
Our main results are based on analyzing the following SDP relaxation \prettyref{sdp:primal}. We construct its dual \prettyref{sdp:dual} (refer to \prettyref{app:lagrangian_calculation} for more details on this dual construction).

\begin{tcolorbox}[]
\begin{multicols}{2}
    \begin{SDP}[Primal]
		\label{sdp:primal}
		\[ \min \sum_{\set{i,j} \in E} 2\inprod{\mathbf{x}_i,\mathbf{x}_j}  \]
		\subjectto
		\begin{align}
			\label{eq:sdp1}
			&\sum_{i \in V}{\norm{\mathbf{x}_i}^2} = 1\\
			\label{eq:sdp2}
			&\norm{\mathbf{x}_i}^2 \leq 1/k & \forall i \in V\\
			\label{eq:sdp3}
			&\inprod{\mathbf{x}_i,\mathbf{x}_j} \leq 0 & \forall \set{i,j} \in E \mper
		\end{align}
		\end{SDP}
		\columnbreak
		\begin{SDP}[Dual]
		\label{sdp:dual}
		\[ \max \,\, \beta - \sum\limits_{i \in V}\gamma_i\]
		\subjectto
		\begin{align}
			\label{eq:sdp5}
			&Y=A -\beta I + k\sum_{i \in V}\gamma_iD_i \nonumber \\
			&\,\,\,\,\,\,\,\,\,\,\,\,\,\,\,\,\,\,\,   \,\,\,\,\,\, +\sum_{\set{i,j} \in E}B_{ij}\paren{\one_{ij} + \one_{ji}} \\
			\label{eq:sdp6}
			&B_{ij} \geq 0, \quad\quad \forall \set{i,j} \in E\\
			\label{eq:sdp7}
			&Y \succeq 0 \mper
		\end{align}
		\end{SDP}
\end{multicols}
\end{tcolorbox}

In \prettyref{sdp:dual}, the Lagrange multipliers  $\beta_i$'s,$\gamma_i$'s and $B_{ij}$'s are our dual variables and $Y$ is the dual SDP matrix. By $\one_{ij}$ we mean an indicator matrix which is one for $(i,j)$ entry and zero elsewhere. 
Similarly, $D_i$ is an indicator matrix which is one for $(i,i)$ entry and zero elsewhere. 
For clarity, we will denote $\sum_{\set{i,j} \in E} B_{ij}\paren{\one_{ij} + \one_{ji}}$ by a matrix $B$.

\paragraph*{Intended solution:}
We denote the primal SDP matrix by $X$ and let $\mathbf{x}_i$ denote the vector corresponding to vertex $i$ such that $X_{ij} = \inprod{\mathbf{x}_i,\mathbf{x}_j}$.

Our intended integral solution to the SDP is $X=\mathbf{g}\mathbf{g}^T$, where $\mathbf{g} \in \mathbb{R}^n$ s.t $g_i=\ffrac{1}{\sqrt{k}}$ for $i \in S_1$, $g_i=-\ffrac{1}{\sqrt{k}}$ for $i \in S_2$ and $0$ otherwise. This solution is obtained by setting,
\begin{equation} \label{eq:integral}
	\mathbf{x}_{i}^* =  
	\begin{cases} 
	\ffrac{\hat{e}}{\sqrt{k}} & \text{if } i \in S_1 \\
	-\ffrac{\hat{e}}{\sqrt{k}} & \text{if } i \in S_2 \\
	0 &  \text{otherwise},
	\end{cases} 
	\end{equation}
where $\hat{e}$ is some unit vector. 

\paragraph*{Weak Duality for fixing dual variables:}
Let $\mathsf{SDPOPT}\paren{G}$ denote the optimal value of the primal SDP; then from the proposed integral solution we have that,
\begin{align*}
    \mathsf{SDPOPT}\paren{G} \leq -2\sum_{\set{i,j} \in E} \inprod{\mathbf{x}_i^*,\mathbf{x}_j^*} = \inprod{A,\mathbf{g}\mathbf{g}^T} = \mathbf{g}^TA\mathbf{g} = -d \mper
\end{align*}
For any feasible solution to the dual \prettyref{sdp:dual}, by weak duality, we know that \begin{align*}
\beta - \sum_{i \in V}\gamma_i \leq \mathsf{SDPOPT}\paren{G} \leq -d\mper
\end{align*}
We note that the upper bound is achievable by setting $\beta=-d$ and $\gamma_i=0,\forall i \in V$.

We will show later that the remaining dual variables $B_{ij}$'s can be chosen in a way that the choice of $\beta=-d$ and $\gamma_i=0,\forall i \in V$ yields a feasible dual solution.

\begin{fact} [Foklore, also see Lemma 2.3 in \cite{LV18}]\label{fact:optimality_conditions}
The primal solution $X=\mathbf{g}\mathbf{g}^T$ is the unique solution to \prettyref{sdp:primal} if there exists a dual matrix $Y$ such that it satisfies constraints in \prettyref{sdp:dual}, with $\beta=-d$ and $\gamma_i=0,\forall i \in V$ and having $\rank(Y)=n-1$ (i.e. $\lambda_2\paren{Y} > 0$).
\end{fact}
\begin{proof}
For completeness, we give a proof in \prettyref{app:sdp_complementary_slackness}.
\end{proof}

\subsubsection{Threshold rank eigenvectors:}
\begin{definition}[Threshold rank of a graph] \label{def:threshold_rank}
For $\tau \in [0,d]$, we define threshold rank of a graph with adjacency  matrix $G$ (denoted by $\rank_{\leq -\tau}(G)$) as,
\begin{align*}
    \rank_{\leq -\tau}(G) = \abs{\set{i : {\lambda_i(G)} \leq - \tau}} \mper
\end{align*}
\end{definition}

We let $P_{-\tau}=\set{\mathbf{v}^{\paren{1}},\mathbf{v}^{\paren{2}},\hdots,\mathbf{v}^{\paren{L_{-\tau}}}}$ (the bottom $L_{-\tau}$ vectors) denote the set of orthonormal eigenvectors of $\restrict{A}{S \times S}$ with eigenvalues smaller than the threshold $-\tau$, breaking ties arbitrarily where $L_{-\tau}=\rank_{\leq -\tau}\paren{\restrict{A}{S \times S}}$. We call these vectors as $\tau$-\textit{threshold rank eigenvectors} of $\restrict{A}{S \times S}$.
Next, we recall a well known fact about the threshold rank of a graph.

\begin{fact}[Folklore]\label{fact:threshold_rank}
$\rank_{\leq -\tau} \paren{\left.A\right|_{S \times S}} \leq \dfrac{kd}{2{\tau}^2} $.
\end{fact}

\begin{proof}
Since, $\restrict{A}{S \times S}$ is the adjacency matrix of a bipartite graph, it's eigenvalue spectrum is symmetric around $0$. Therefore the number of eigenvalues with absolute value greater then or equal to $\tau$ is given by $2\, \rank_{\leq -\tau}\paren{\restrict{A}{S \times S}}$ and are bounded as,
\begin{align*} 
2   {\tau}^2 \rank_{\leq -\tau}(\restrict{A}{S \times S})  \leq  \sum_{i}{\lambda_i^2(\restrict{A}{S \times S})} \leq \norm{\restrict{A}{S \times S}}^2_F =kd.
\end{align*}
\end{proof}

\noindent
We note that a similar notion of threshold rank has appeared in other works \cite{ABS10,AG11,BRS11,GT13} etc. 

\subsubsection{Spectral embedding vectors:}

\begin{definition}[Spectral embedding vectors]
Given the planted bipartite graph $S$ and the matrix of bottom $L_{-\tau}$ orthonormal eigenvectors $W^T_{-\tau}= 
\begin{bmatrix}
\mathbf{v}^{\paren{1}} & \mathbf{v}^{\paren{2}} & \hdots & \mathbf{v}^{\paren{L_{-\tau}}}
\end{bmatrix}
$, we define the spectral embedding of a vertex $i \in S$ as the $L_{-\tau}$-dimensional vector given by $\mathbf{w}^{\paren{i}} = W_{-\tau}\mathbf{e}_i$ where $\mathbf{e}_i$ is a vector with one in the $i^{th}$ coordinate and zero elsewhere.
\end{definition}

Informally, these are the vectors obtained by looking at the subspace of the columns of $W^T_{-\tau}$ where the vertex $i$ is mapped to the $i^{th}$ column of $W^T_{-\tau}$. These spectral embedding vectors have been explored in various works on graph partitioning as \cite{NJW01,LOT12,LRTV12} etc. It is known that these spectral embedding vectors are \say{well spread}, formally referred to as being in an isotropic\footnote{Typically, isotropicity is a property of distribution. We say a distribution is isotropic if the mean of a random variable sampled from the distribution is zero and it's covariance matrix is an identity matrix.} position. We define these set of vectors to be in an isotropic position if $\sum_{i \in S}\mathbf{w}^{(i)}=0$ and $\sum_{i \in S}\mathbf{w}^{(i)}{\mathbf{w}{^{(i)}}^T}=I$ where $I$ is an $L_{-\tau} \times L_{-\tau}$ sized identity matrix. The condition that $\sum_{i \in S}\mathbf{w}^{(i)}{\mathbf{w}{^{(i)}}^T}=I$ can equivalently be written as $\sum_{i \in S}\inprod{\mathbf{y},\mathbf{w}^{(i)}}^2=1,\forall \mathbf{y} \text{ with }\norm{\mathbf{y}}_2=1$.

\begin{lemma}[Folklore]
The spectral embedding vectors are in an isotropic position.
\end{lemma}
\noindent
For a proof,
we refer the reader to the work \cite{LRTV11}.

\subsection{Proof Overview} \label{sec:proof_overview}
For the sake of simplicity, we will assume that the graph is sampled as per the random planted model (\prettyref{def:planted_model}). We will also allow an action of a monotone adversary (as in step 5) on this model; but we analyze its action separately (in \prettyref{sec:adversary_proof_overview}). The main ideas for the semi-random model (\prettyref{def:semi-random_model}) are essentially the same, and the additional steps to handle them is just a technical adjustment. 
\subsubsection{Spectral Approaches}
We start with some natural spectral approaches for recovering the planted set. These approaches have found some success, e.g. in recovering planted cliques/independent sets, planted bisection, planted $k$-colorable graphs (refer work \cite{McS01} for details). We recall from our earlier discussion, that the interesting regimes for this problem are $k=\Omega\paren{\sqrt{n}}$ and $d \approx pk$.

\paragraph*{Detecting planted bipartitions and why it is easy:}  
We note that the \textit{detection problem} i.e. detecting the presence of bipartite graph as constructed in the random planted model (\prettyref{def:planted_model}) against the null hypothesis of \Erdos-\Renyi graph $G_{n,p}$, is easy when $k=\Omega\paren{\sqrt{n}}$. Formally one notes that given two distributions
\begin{align*}
    H_0: G \sim G\paren{n,p} \text{ against } H_1: G \sim G\paren{n,k,d,p} \text{ as per \prettyref{def:planted_model}},
\end{align*}
the spectral test, which outputs $H_1$ when $\lambda_{1}\paren{G} \leq -d$ and $H_0$ otherwise, is correct almost surely for $d \approx pk$ and $k \geq \ffrac{c\sqrt{n}}{p}$ where $c>0$ is a large enough constant. This is because for a $G_{n,p}$ graph, the smallest eigenvalue is greater than $-2\sqrt{n}$ almost surely (\prettyref{claim:random_matrix_norm}), while for a graph with planted bipartite subgraph, the smallest eigenvalue is smaller than $-d$ since the vector $\one_{S_1}-\one_{S_2}$ already achieves Rayleigh Quotient of value $-d$.

\paragraph*{The challenges in exact recovery:} However, as expected, the exact recovery problem is more challenging. There are some works that look at these planted problems on an individual basis (\cite{Bop87},\cite{AKS98}). They typically rely on the spectral bounds of perturbation matrices and the framework of Davis-Kahan theorem (refer \cite{Ver18}) to identify eigenvector(s) indicating the planted set. However, we need a sufficient eigengap\footnote{Typically around the bottom eigenvector(s) or the top eigenvector(s).} to apply these results from perturbation theory. Since our planted graph $S$ in the random planted model is an arbitrary bipartite graph, it can have any number of eigenvalues close to the smallest eigenvalue $-d$ and hence we may not have such an eigengap.

A unified spectral framework for random planted models was given by McSherry \cite{McS01} (further refined in the work \cite{Vu18}). Here, one can check that we cannot satisfy the conditions in Observation 11 of this work \cite{McS01} if the planted set has size $o(n)$. Again, the reason is because the planted bipartite graph is arbitrary. Since the planted bipartite graph can have arbitrary rank we cannot get the constants $\gamma_1$ in \cite{McS01} to be small enough to recover in $o(n)$ regimes. It is also easy to verify that this framework works if the planted bipartite graph is also a random graph, for regimes of $k=\Omega\paren{\sqrt{n}}$  and $ d\approx pk$ (say by choosing edge probability for the random planted bipartite graph as $p'=2p$).

\paragraph*{A subspace enumeration style approach:}
Another spectral approach, inspired from the works of  \cite{KT07,ABS10,Kol11,KLT17} is to apply the \textit{subspace enumeration} technique to recover a large fraction of planted set $S$. 
Here we first identify (in \prettyref{lem:close_vector_in_span}) that the vector $\mathbf{u}=\one_{S_1}-\one_{S_2}$ has a large projection on the space spanned by $\tau'$-threshold rank eigenvectors of $A$ (for choice of $\tau'=d/2$). Note that this vector $\mathbf{u}$ identifies the planted set (as well as the planted bipartition), and therefore we call it the \textit{signed indicator vector}.
We then do a standard $\varepsilon$-net construction to find a vector $\mathbf{y}$ close to $\mathbf{u}$ and use $\mathbf{y}$ to recover a large fraction of planted set $S$ (\prettyref{lem:epsilon_net_construction} and \prettyref{lem:large_fraction_recover}).
We can recover the remaining set of vertices by an argument due to the work \cite{GLR18} (\prettyref{lem:matching_argument}), where they distinguish vertices by the size of matching in induced neighborhoods. Putting all this together, we can prove \prettyref{thm:subspace_informal}.

The running time of the procedure described above is exponential in $L_{-\tau}$  where $L_{-\tau}=\bigO\paren{\ffrac{1}{p}}$ for $\tau=\Omega\paren{d}$ (follows from Lemma \prettyref{fact:threshold_rank}). 
Therefore, for many special classes of instances such as, (i) when the probability $p=\Omega(1)$ and $d=\Omega(pk)$, (ii) when the planted graph is a complete bipartite graph (this is the balanced biclique problem) and $d = \Omega(pk)$, (iii) when the planted bipartite graph is $d$-regular random graph for $d = \Omega(pk)$ or (iv) more generally when the planted graph is a $d$-regular expander graph for $d = \Omega(pk)$ we have $L_{-\tau}= \bigO\paren{1}$ and this already gives us a polynomial-time algorithm.

However, as stated earlier, we want to solve the problem in $p=o(1)$ regimes. To accomplish this, we shift our focus to the SDP formulation we mentioned in \prettyref{sdp:primal}. Also, for other problems in this literature (planted clique, planted bisection, planted $k$-colorable, Stochastic Block models etc), only SDP's have provable guarantees of working in the presence of such monotone adversaries (refer Chapter 10 in \cite{Rou21} for more intuition on this). 

\subsubsection{Traditional SDP Analysis}
\label{sec:traditional_sdp}
Now we overview our SDP-based approach to solving the problem. We will see that the difficulties in the spectral approach will translate to showing the feasibility of the dual SDP solution. However, we have more freedom here since we have the dual variables to work with and we can use them and try to enforce the optimality of the dual solution.

\paragraph*{Characterizing dual variables through optimality conditions:} 
A standard technique for analyzing SDP relaxations (like our \prettyref{sdp:primal}) is to show optimality by constructing a dual solution that matches the $\SDPOPT\paren{G}$ value of the primal in a manner that the dual matrix $Y$ is positive semi-definite and has rank $n-1$, (see \prettyref{fact:optimality_conditions}). 

These impose a \say{wish list} of desired conditions, which can be used to characterize our dual variables
\begin{multicols}{2}\setlength{\columnseprule}{0pt}
\begin{enumerate}
\item $\beta=-d$ (Optimal objective)
\item $\inprod{\mathbf{g}\mathbf{g}^T,Y} =0$ (Complementary slackness)
\item $Y\succeq 0$ (Dual feasibility)
\item $\lambda_{2}\paren{Y} > 0 $ (Strong duality).
\end{enumerate}
\end{multicols}

Using weak duality we set $\beta=-d$ and $\gamma_i=0,\forall i \in V$ to match the optimal primal objective value of $\SDPOPT=-d$. We expand upon the complementary slackness condition as,
\begin{align*}
    \inprod{\mathbf{g}\mathbf{g}^T,Y} = \mathbf{g}^TY\mathbf{g} = \mathbf{g}^T\paren{A+dI}\mathbf{g} + \mathbf{g}^TB\mathbf{g} = 0 + \mathbf{g}^TB\mathbf{g} = \mathbf{g}^TB\mathbf{g} \mper
\end{align*}
Therefore the complementary slackness condition gives us that,
\begin{align} \label{eq:duality_dual_variables_condition}
    \sum_{i \in S}\sum_{\substack{j\in S\\\set{i,j}\in E}}B_{ij}=0 
\end{align}
and since the SDP dual requires that $B_{ij} \geq 0$, it implies that $B_{ij} = 0$ for all $(i,j) \in E(S_1,S_2)$.
Now using the characterization of dual variables from conditions (1) and (2), one tries to show the feasibility of the dual and the strong duality rank condition. Typically, this characterization turns out to be rather weak. So, we refer to the dual variables set so far (ensuring condition (1) and (2) are satisfied) as \emph{weakly characterized}.

\paragraph*{Showing optimality of dual solution through weakly characterized dual variables }
For certain problems in semi-random models, such as the planted clique problem \cite{FK00}, community detection in SBM \cite{ABH16}, the weak characterization above suffices. We are able to show that the weakly characterized dual solution satisfies conditions (3) and (4). This is typically done by invoking some standard results for random matrix bounds and concentration inequalities. 
In our setup, satisfying condition (3) requires that
\begin{equation}\label{eq:simple_conditions_dual}
    \lambda_{\min}\paren{Y} \geq 0 \text{ which is implied if } \lambda_{\min}\paren{A}+d+\lambda_{\min}\paren{B} = \lambda_{\min}(A)+d \geq 0 \mper 
\end{equation}
However, in our random planted model, the smallest eigenvalue of $A$ can be smaller than $-d-2\sqrt{pn}$ and condition (3) may not hold (as per choice of $B_{ij}$'s dictated from equation \prettyref{eq:duality_dual_variables_condition}).
Thus we need a stronger characterization of dual variables to satisfy the conditions (3) and (4). In our problem, we need to make use of the large number of unused dual variables $B_{ij}'s$ for $\set{i,j} \in E \cap \set{\paren{V \times V} \setminus \paren{S \times S}}$

\paragraph*{Guessing/Constructing the dual certificate}

Now we discuss an approach of making the dual matrix satisfy conditions (3) and (4) by guessing the dual variables thus giving an explicit setting of dual variables. This is typically done by assigning some sort of meaning to dual variables and guessing their values based on the input instance. 
This approach has found reasonable success in other recovery problems like the planted bisection problem \cite{FK01}, coloring semi-random graphs \cite{CO07}, decoding binary node labels from censored edge measurements  \cite{ABBS14}, and planted sparse vertex cuts \cite{LV18}.

Therefore, we may expect to guess a nice setting of dual variables that satisfy equation \prettyref{eq:simple_conditions_dual}. However, if one takes a deeper look at this approach, the task again reduces to applying results from perturbation theory. Again, such an approach would work if the planted bipartite graph were also a random graph or an expander, since there would only be a single eigenvector whose corresponding eigenvalue disobeys equation \prettyref{eq:simple_conditions_dual}, and one could choose the dual variables constructively to handle it and make it satisfy condition (3).

However, for an arbitrary planted bipartite graph, we can have a lot of eigenvalues in the interval $[-d,-d-2\sqrt{n}]$ (and hence the entire graph $A$ can have a lot of eigenvalues in the interval  $[-d-2\sqrt{n},-d]$).
Therefore, we need a more principled approach to deal with the corresponding eigenvectors of the planted graph having eigenvalue close to $-d$ (as we pointed out earlier, there can be $\bigO\paren{\ffrac{1}{p}}$ such eigenvectors).

\subsubsection{Calibrating the eigenvectors}
Now we present our approach towards satisfying conditions (3) and (4), which is to \emph{calibrate the eigenvectors}. We will see that, this calibration will further complicate our requirements on the dual variables, however we will argue in \prettyref{sec:proof_overview_pseudorandom} on how we manage that.

\paragraph*{Obtaining optimality of Primal SDP by assuming existence of a certifying $B$:} 
It is now clear that our Achilles' heel are the eigenvalues (and corresponding threshold rank eigenvectors\footnote{For an appropriate choice of $\tau $, which we decide later, these will be the threshold rank eigenvectors.})  of the planted graph in the interval $[-d,-d-2\sqrt{n}]$. If we were allowed to ignore these vectors it's easy to see that equation \prettyref{eq:simple_conditions_dual} and hence condition (3) holds.

Our core idea is to extend (by padding with $0$'s so that they are the right length) the threshold rank eigenvectors of $\restrict{A}{S \times S}$ to be the eigenvectors of the dual matrix $Y$. 
Recall, the eigenvalues of $\restrict{A}{S \times S}$ lie in the interval $\Brac{-d,d}$. Now take a threshold rank eigenvector of $\restrict{A}{S \times S}$ (say with eigenvalue $\lambda_l$). We wish to calibrate such a threshold rank eigenvector to be an eigenvectors of $Y$ with eigenvalue $d+\lambda_l$.
If we are able to achieve this calibration, we need not bother about the $2\sqrt{n}$ term since now these eigenvectors have a non-negative quadratic form\footnote{The quadratic form of vector $\mathbf{x}$ with a matrix $Y$ is a number given by $\mathbf{x}^TY\mathbf{x}$}.

The only thing at our disposal for this calibration are the unused (so far) dual variables $B_{ij}$'s. Denote this set of threshold rank eigenvectors of $\restrict{A}{S \times S}$ as $P_{-\tau}$. Given this set $P_{-\tau}$, achieving this calibration can be expressed as satisfying the system of equations, 

\begin{align} 
\sum_{i \in S}{\mathbf{v}^{\paren{l}}_i\paren{A_{ij}+B_{ij}}} = 0, \quad \forall j \in V \setminus S, \forall\, \mathbf{v}^{\paren{l}} \in P_{-\tau} \mper \label{eq:eigenvector_row_proof_overview} 
\end{align}

\noindent
\begin{itemize}
    \item $L$ different system of equations $\cE_l$, one for each $\mathbf{v}^{\paren{l}}$.
    \item Each system $\cE_l$ involves $|S| \times |V \setminus S|$ variables $B_{ij}$ where $i \in S$ and $j \in V \setminus S$. 
\end{itemize}

\noindent
Now, all we need is a setting of $B_{ij}$'s such that the system of equations is satisfied. However, this will still not be enough. We note that if this system of equations were to have a solution, we would have set some of these $B_{ij}$ variables to non-zero values. Therefore our equation \prettyref{eq:simple_conditions_dual} would now need to be modified to showing that $\lambda_{\min}\paren{A}+d+\lambda_{\min}\paren{B} \geq 0$.

The way we deal with this is by noting that we can tune $\tau>0$ apriori to be sufficiently large for this calibration such that $\lambda_{\min}(A)+d \geq \eta$ where $\eta >0$; and now impose an additional constraint on the matrix of dual variables that $\norm{B} \leq \eta$. At this point, it seems highly suspicious as whether such $B_{ij}$'s exist. However, if we table these considerations aside and for choice of $\tau=2d/3$ and $\eta = \tilde\bigO\paren{pk}$ we can indeed show that condition (3) and condition (4) of our \say{wish list} are met (\prettyref{lem:dual_matrix_psd}) and we get the desired integral primal solution.

\subsubsection{Setting of dual variables} \label{sec:proof_overview_pseudorandom}
In this section we show that there exists a  matrix of non-negative dual variables $B_{ij}$'s that satisfies the system of equations \prettyref{eq:eigenvector_row_proof_overview} and  $\norm{B}_2 =\tilde\bigO\paren{pk}$.

\paragraph*{An LP formulation and Farkas Lemma based approach.}

We start by observing that the condition $\norm{B}_2 = \tilde\bigO\paren{pk}$ is  implied by a condition that $B_{ij} \leq t = \bigO_{p}\paren{\sqrt{\ffrac{\log k}{k}}},\forall (i,j) \in \paren{S \times \paren{V \setminus S}} \cup \paren{\paren{V \setminus S} \times S}$ (\prettyref{cor:entrywise_bound}).
Also, since these system of equations \prettyref{eq:eigenvector_row_proof_overview} only concerns the non-negative dual variables $B_{ij}$'s with $\set{i,j} \in \paren{S \times \paren{V \setminus S}} \cup \paren{\paren{V \setminus S} \times S}$, we set the rest of them to $0$. 

We now reorganize our collection of linear systems in \prettyref{eq:eigenvector_row_proof_overview} as follows. 
\begin{itemize}
    \item For $j \in V \setminus S$, define a system of equations $\cF_j$. 
    \item In all, this gives a collection of systems $\{\cF_j\}_{j \in V \setminus S}$. Each system contains $L_{-\tau} \times |S|$ variables. In particular, the system $\cF_j$ is expressed in the standard form
    $W_{-\tau}\mathbf{x}=\mathbf{b}$, where $W_{-\tau} \in \R^{L_{-\tau} \times k}$ is a matrix formed by stacking the vectors $\mathbf{v}^{\paren{l}} \in P_{-\tau}$ as rows.
\end{itemize}

Fix $j \in V \setminus S$ and consider the system $\cF_j$. The vector $\mathbf{b}$ in this system is a row vector of size $L_{-\tau} \times 1$ and has entries given by $b_l = - \sum_{i \in S}{A_{ij}v^{\paren{l}}_i}, \forall l \in [L_{-\tau}]$ and  $\mathbf{x}$ here is a row vector of size $k \times 1$ where the entry $x_i = B_{ij}$ (recall that we have fixed a $j \in V \setminus S$). However since $B_{ij}$'s are not arbitrary variables but dual variables of \prettyref{sdp:dual}, these are required to be non-negative and should only be defined for $i \in N(j)$. Since the graph on $S \times \paren{V \setminus S}$ is random, the choice of random edges while choosing $N(j)$ (in model construction) corresponds to setting those $B_{ij}=0$ whenever the edge is not chosen.
Let $\tilde W_{-\tau}$ denote the submatrix after removing the columns corresponding to $i \notin N(j)$ and recall $t$ is our upper bound on the entries of $B$ matrix as mentioned above. We then consider the following feasibility LP formulation for this problem of finding appropriate $B_{ij}$'s.
\begin{tcolorbox}
		\begin{LP}
		\label{lp:primal_overview}
		\begin{align}
			\label{eq:lp1}
			&\tilde W_{-\tau}\mathbf{x}=\mathbf{b}\\
			\label{eq:lp2}
			&0 \leq \mathbf{x} \leq t\one \mper
		\end{align}
		\end{LP}
\end{tcolorbox}

For simplicity, consider the case $p=1$, i.e.  when $A_{ij} = 1$ for all $i \in S, j \not\in S$. Then we have that for any vector $\mathbf{y} \in \mathbb{R}^{L_{-\tau}}$,

\begin{align}\label{eq:expand_bty}
    \mathbf{b}^T\mathbf{y} 
    &= \sum_{r\in [L_{-\tau}]}b_ry_r = -\sum_{r \in [L_{-\tau}]}\sum_{i \in S}v^{\paren{r}}_iy_r = -\sum_{i \in S}\sum_{r \in [L_{-\tau}]}v^{\paren{r}}_iy_r\\ 
    &=  -\sum_{i \in S}\sum_{r \in [L_{-\tau}]}w^{\paren{i}}_ry_r = -\sum_{i \in S}\inprod{\mathbf{w}^{\paren{i}},y} = -\inprod{\sum_{i \in S}{\mathbf{w}^{\paren{i}}},\mathbf{y}}=0
    \mper
\end{align}
Using the standard variant of Farkas' Lemma, this immediately implies the existence of a solution to equation \prettyref{eq:eigenvector_row_proof_overview}. However, in general, for $p <1$, we need to do more work here.

We apply a more general version of Farkas' Lemma (\prettyref{cor:farkas_primal_feasible}), and we have that satisfying this LP in the general case corresponds to showing that for some $t>0$, the following holds.
\begin{align} \label{eq:farkas_condition_proof_overview}
    \forall\; \mathbf{y} \in \mathbb{R}^{L_{-\tau}},\forall \; \mathbf{z}\geq 0, \; \tilde W_{-\tau}^T\mathbf{y} + \mathbf{z} \geq 0 \implies \mathbf{b}^T\mathbf{y}+t\inprod{\mathbf{z},\one} \geq 0 \mper
\end{align}
The first term in the expression, $\mathbf{b}^T\mathbf{y}$, can be expanded as in equation \prettyref{eq:expand_bty} to obtain
\begin{align*}
    \mathbf{b}^T\mathbf{y}=-\sum_{i \in N(j)}\inprod{\mathbf{w}^{\paren{i}},\mathbf{y}} \text{ and using } 
    \paren{\tilde W_{-\tau}^T\mathbf{y}}_i = \inprod{\mathbf{w}^{\paren{i}},\mathbf{y}}  \text{ we have } z_i \geq -\inprod{\mathbf{w}^{\paren{i}},\mathbf{y}}.
\end{align*}
We give a proof by contradiction (\prettyref{prop:farkas_apply}) for equation \prettyref{eq:farkas_condition_proof_overview}.
By contradiction there exists a $\mathbf{y}$ and a $\mathbf{z}$ such that $\mathbf{b}^T\mathbf{y}+t\inprod{\mathbf{z},\one} < 0$. 
We choose $\mathbf{z}' \leq \mathbf{z}$ by setting $z'_i=\max\set{0,-\inprod{\mathbf{w}^{(i)}},\mathbf{y}}$ and argue that it is enough to show contradiction for $t,\mathbf{y}$ and $\mathbf{z}'$.
Using the expressions for $\mathbf{b}^T\mathbf{y}$ as above, this translates to showing that
\begin{align} \label{eq:farkas_no_solution_proof_overview}
    \sum_{i \in N(j)}\inprod{\mathbf{w}^{\paren{i}},\mathbf{y}} + t\sum_{i\in N(j)}{\min\set{0,{\inprod{\mathbf{w}^{\paren{i}},\mathbf{y}}}}} >0,
\end{align}
has no solution. We show that equation \prettyref{eq:farkas_no_solution_proof_overview} does not hold for our desired choice of $t=\bigO_{p}\paren{\sqrt{\ffrac{\log k}{k}}}$. 

We note that the second term in equation \prettyref{eq:farkas_no_solution_proof_overview} is $\geq 0$, and we wish the inequality to not hold for as small a value of $t$ as possible; therefore, we seek an upper bound on both terms. 

\paragraph*{Structure of threshold rank/spectral embedding vectors.}
To upper bound the first term, it might be helpful to understand the structure of the spectral embedding vectors $\mathbf{w}^{\paren{i}}$. Since these are intimately connected to the threshold rank eigenvectors $\mathbf{v}^{\paren{l}}$, we use these eigenvectors to characterize them. For convenience we let $\mathbf{v}^{\paren{l}} \in P_{-\tau}$ have unit norm, then in \prettyref{lem:eigenvec_l_inf_norm} we show that $\norm{\mathbf{v}^{\paren{l}}}_{\infty}=\norm{\mathbf{w}^{\paren{i}}}_{\infty} \leq \ffrac{2}{\sqrt{d}}$. Since $i \in N(j)$ are sampled randomly, we can  use the Hoeffding bounds to upper bound the $l_{\infty}$ norm for $\sum_{i \in N(j)}\mathbf{w}^{\paren{i}}$ and hence upper bound our first term (\prettyref{lem:sum_spectral_embed_bound}) by $\bigO_{p}\paren{\sqrt{\log k}}$ with high probability. 
We choose our parameters such that the vectors in $P_{-\tau}$ are orthogonal to $\one_S$. For the embedding vectors, this translates to saying that $\sum_{i \in S}\mathbf{w}^{\paren{i}}=0$.

Towards bounding the second term, we use these \textit{spectral embedding vectors}. The spectral embedding vectors are isotropic for $p=1$ (already where we can easily show that  equation \prettyref{eq:farkas_no_solution_proof_overview} does not hold and we are done).
However, for $p<1$, we have $i \in N(j)$ (and corresponding embedding vectors) being sampled randomly as per $G_{n,p}$ distribution.
Here, in \prettyref{lem:matrix_concentration_bound}, we show that by using Matrix Bernstein concentration we can get close to isotropic vectors,
\begin{align}\label{eq:matrix_conc_proof_overview}
    \sum_{i \in N(j)}\inprod{\mathbf{y},\mathbf{w}^{\paren{i}}}^2 \geq \ffrac{p}{2} && \paren{\text{This shows that embedding vectors are $\ffrac{p}{2}$-isotropic
    }.}
\end{align}

\paragraph*{Showing existence of a solution to \prettyref{lp:primal_overview}.}
Now, we look at two cases; the first case where the negative terms dominate the summand in equation \prettyref{eq:farkas_no_solution_proof_overview}, then we use the eigenvector structure that $\norm{\mathbf{w}^{\paren{i}}} \leq \ffrac{2}{\sqrt{d}}$ and we are done; for the other case where the positive terms dominate, we relate the positive terms to the negative terms again using the bound we obtained from the eigenvector structure $\norm{\sum_{i \in N(j)}\mathbf{w}^{\paren{i}}} =\bigO_{p}\paren{\sqrt{\log k}}$. 

Therefore, we argue in \prettyref{lem:farkas_condition_second_term}, that we can upper bound the second term by $-\Omega_{p}\paren{\sqrt{k}}$.
Therefore using these bounds we show in \prettyref{prop:farkas_apply}, that for a choice of $t$ as we obtained above of $t=\bigO_{p}{\paren{\sqrt{\ffrac{\log k}{k}}}}$, equation \prettyref{eq:farkas_no_solution_proof_overview} does not hold. As discussed earlier this implies that the there exists a dual such that conditions (1)-(4), equation \prettyref{eq:eigenvector_row_proof_overview} and $\norm{B}_2=\tilde\bigO\paren{pk}$ holds which further implies that the primal SDP is feasible. Further, if the graph is connected; the  \textit{signed indicator vector}  would be the only eigenvector with eigenvalue $-d$ (after padding to make these the eigenvectors of $Y$), this would be the only eigenvector of $Y$ with eigenvalue $0$. Using \prettyref{fact:optimality_conditions}, this implies that the proposed integral solution in equation \prettyref{eq:integral} is the only integral solution and hence Cholesky Decomposition of our SDP matrix returns the \textit{signed indicator vector} and thus our planted set.

\subsubsection{Low degree regimes} \label{sec:low_degree_regimes}
The discussion above about SDP holds only where $d = \gamma pk $ where $\gamma \geq 2/3$. Note that this covers our interesting regimes when $d \approx pk$ where the problem is non-trivial. The other case where $d \leq \ffrac{2pk}{3}$, can actually be trivially solved for $k=\Omega_{p}\paren{\sqrt{n \log n}}$ using a degree counting argument along the lines of \cite{Kuc95} as we discuss below.

Now we consider the regimes when $d = \gamma pk$ with $\gamma \leq \ffrac{2}{3}$. We show that a simple algorithm that collects the bottom $k$ degrees of the graph will work in these regimes since the vertices in $S$ will have smaller degrees compared to vertices in $V \setminus S$.

\RestyleAlgo{ruled}
\begin{minipage}{\linewidth}
\hspace{-1.75em}
	\begin{algorithm}[H]
		\caption{}
		\label{alg:three}
		\begin{algorithmic}[1]
		\setlength{\algomargin}{1pt}
			\REQUIRE $G=(V, E)$, sampled as per \prettyref{def:planted_model} with adversary as per \prettyref{step:adversary}.
			\ENSURE The set of vertices in planted bipartite graph (with high probability) ${S}$.
			\STATE Sort the degrees of the vertices in $G$.
			\STATE Return $\mathcal{S}$ be the set of bottom $k$ degrees after sorting..
		\end{algorithmic}
	\end{algorithm}
\end{minipage}

\begin{lemma} \label{lem:low_degree_proof}
For $k \geq 6\sqrt{\ffrac{6n\log n}{p}}$, \prettyref{alg:three} returns the planted set $S$ with high probability (over the randomness of the input).
\end{lemma}
\begin{proof}
For a vertex $v \in S$ the expected degree is $d+p(n-k)$. We note that this is smaller than $pn$ since $d \leq 2pk/3$. We can upper bound the degree of $v$ (denoted $d(v)$), with high probability (over the randomness of the input) using \prettyref{fact:chernoff_upper} as,
\begin{align*}
    \Pr{d(v) \geq d + p(n-k)+\sqrt{6pn\log n}} \leq  \exp\paren{\frac{-pn(\sqrt{6\log n}/\sqrt{pn})^2}{3}}
    =\frac{1}{n^2}.
\end{align*}
Using a union bound over all $v \in S$, we have that for an any $v \in S$,
\begin{align*}
    \Pr{d(v) \geq d + p(n-k)+\sqrt{6pn\log n}} \leq  \frac{1}{n}.
\end{align*}
Therefore we have with high probability (over the randomness of the input) that $d(v) \leq d+p(n-k) + \sqrt{6pn \log n}$. Similarly for a vertex $v' \notin S$ the degree can be lower bounded with high probability (over the randomness of the input) using \prettyref{fact:chernoff_lower} as,
\begin{align*}
    \Pr{d(v') \leq pn - \sqrt{6pn\log n}} \leq  \exp\paren{\frac{-pn(\sqrt{6\log n}/\sqrt{pn})^2}{2}}=\exp\paren{-3\log n}=\frac{1}{n^3}.
\end{align*}
Now using a union bound over all $ v'\notin S$, we have with high probability (over the randomness of the input) that $d(v') \geq pn - \sqrt{6pn \log n}$. Therefore, with high probability (over the randomness of the input), the degrees differ by, 
\begin{align} \label{eq:degree_concentrate}
    d(v')-d(v) \geq pk-d - 2\sqrt{6pn\log n} \geq \dfrac{pk}{3} - 2\sqrt{6n\log n} \mper
\end{align}
where we have used $d \leq 2pk/3$.
It is also evident from equation \prettyref{eq:degree_concentrate} that for $k \geq 6\sqrt{\ffrac{6n\log n}{p}}$, with high probability (over the randomness of the input), the degree for a vertex $v \in S$ is smaller than degree of any vertex $v' \notin S$.
\end{proof}

\subsubsection{Action of Adversary}
\label{sec:adversary_proof_overview}
Finally, the action of adversary (allowed to add edges in $\paren{V \setminus S} \times \paren{V \setminus S}$ for $d \geq 2pk/3$) is discussed in \prettyref{sec:adversary_action}.
We show that the inductive argument given by \cite{FK01} also works for our case.
This argument also extends to the semi-random model in \prettyref{def:threshold_rank}.

For $d \leq 2pk/3$ regimes, \prettyref{alg:three} continues to return the planted set, since the action of adversary only amplifies the difference of degree for a vertex $v \in S$ and vertices $v' \notin S$.
This argument does not extend to the semi-random model in \prettyref{def:threshold_rank}.

\section{Exact recovery in polynomial time using SDP} \label{sec:sdp_section}
In this section, we consider the problem of recovering the planted bipartite graph constructed as per our semi-random model in  \prettyref{def:semi-random_model}. It is crucial to emphasize that all our proofs in \prettyref{sec:optimal_dual_construction} and in \prettyref{sec:dual_feasible} consider a graph sampled from \prettyref{def:semi-random_model} \emph{before the action of the adversary}. Handling the action of the adversary is done via standard arguments in \prettyref{sec:adversary_action}. In particular, we overload \prettyref{def:semi-random_model} in \prettyref{sec:optimal_dual_construction} and \prettyref{sec:dual_feasible}, and sampling a graph distributed according to \prettyref{def:semi-random_model} refers to obtaining a graph from this model before step 5 (the adversary action step).

\begin{theorem}[Formal version of \prettyref{thm:arbitrary_informal}] \label{thm:arbitrary_formal}
For $n,k,d,p$ satisfying $k \geq \dfrac{512}{\alpha p^{\ffrac{7}{2}}}\sqrt{n \log n}$ and $p \geq 5\paren{\ffrac{\paren{\log k}}{\gamma^4 k}}^{\ffrac{1}{6}}$ and $d=\gamma pk$ for $\gamma \geq 2/3$ and $0 \leq \alpha \leq \gamma -(1/2)$, there exists a deterministic algorithm which recovers the planted set $S$ in an instance generated as per \prettyref{def:semi-random_model}, exactly with high probability (over the randomness of the input).
\end{theorem}

We did not make any attempt to optimize the constants above (in \prettyref{thm:arbitrary_formal}) and the specific values we use are a result of choices we make for ease of calculation.

\subsection{Constructing an optimal dual}
\label{sec:optimal_dual_construction}

In this section we show how to construct an optimal dual to the SDP.
The main workhorse of our algorithm in high degree regimes is our \prettyref{sdp:primal}. By high degree regimes we mean that $\gamma \geq \ffrac{1}{2}+ \alpha$ where $d=\gamma pk$. We recall that $\alpha$ is a small positive constant smaller than $\ffrac{1}{6}$. We repeat the SDP here for reader's convenience.

\begin{tcolorbox}[]
\begin{multicols}{2}
    \begin{SDP}[Primal]
		\label{sdp:primal_repeat}
		\[ \min \sum_{\set{i,j} \in E} 2\inprod{\mathbf{x}_i,\mathbf{x}_j}  \]
		\subjectto
		\begin{align}
			\label{eq:sdp11}
			&\sum_{i \in V}{\norm{\mathbf{x}_i}^2} = 1\\
			\label{eq:sdp21}
			&\norm{\mathbf{x}_i}^2 \leq 1/k & \forall i \in V\\
			\label{eq:sdp31}
			&\inprod{\mathbf{x}_i,\mathbf{x}_j} \leq 0 & \forall \set{i,j} \in E \mper
		\end{align}
		\end{SDP}
		\columnbreak
		\begin{SDP}[Dual]
		\label{sdp:dual_repeat}
		\[ \max \,\, \beta - \sum\limits_{i \in V}\gamma_i\]
		\subjectto
		\begin{align}
			\label{eq:sdp51}
			&Y=A -\beta I + k\sum_{i \in V}\gamma_iD_i \nonumber \\
			&\,\,\,\,\,\,\,\,\,\,\,\,\,\,\,\,\,\,\,   \,\,\,\,\,\, +\sum_{\set{i,j} \in E}B_{ij}\paren{\one_{ij} + \one_{ji}} \\
			\label{eq:sdp61}
			&B_{ij} \geq 0, \quad\quad \forall \set{i,j} \in E\\
			\label{eq:sdp71}
			&Y \succeq 0 \mper
		\end{align}
		\end{SDP}
\end{multicols}
\end{tcolorbox}

\begin{corollary}\label{cor:threshold_rank}
For the planted bipartite graph on set $S$, we can bound it's $\tau$-threshold rank as $\rank_{\leq -\tau} \paren{\left.A\right|_{S \times S}} \leq \dfrac{\gamma pk^2}{{2\tau}^2} $.
\end{corollary}

\begin{proof}
The result follows by setting $d=\gamma pk$ in \prettyref{fact:threshold_rank}.
\end{proof}

\noindent
Using  this threshold rank, we earlier defined the corresponding eigenvectors as \textit{threshold rank eigenvectors}.
We recall that our main idea is to show that the dual variables $B_{ij}$'s can be chosen such that a $\tau$-\textit{threshold rank eigenvector} $\mathbf{v}^{\paren{l}}$ with eigenvalue $\lambda_l$ of $\restrict{A}{S \times S}$ can be extended (by padding it with $0's$) to be the eigenvector\footnote{With slight abuse of notation we will call these padded vectors of length $n$ also as $\mathbf{v}^{\paren{l}}$.} of the dual matrix $Y$ with eigenvalue $d+\lambda_l$. For this to hold true we additionally require (in addition to the constraints (1)-(4) in \prettyref{sec:traditional_sdp}), that we have,
\begin{align} 
\sum_{i \in S}{\mathbf{v}^{\paren{l}}_i\paren{A_{ij}+B_{ij}}} = 0, \quad \forall j \in V \setminus S, \forall\, \mathbf{v}^{(l)} \in P_{-\tau} \mper \label{eq:eigenvector_row} 
\end{align}
Additionally, for reasons that will be evident in the proof of \prettyref{lem:dual_matrix_psd}, we require that the dual matrix $B$ satisfies  $\norm{B} \leq \ffrac{896\sqrt{n\paren{\log k}}}{p^{\ffrac{5}{2}}}$.
We defer proving the existence of such $B_{ij}$'s to \prettyref{sec:dual_feasible}. However since the equation \prettyref{eq:eigenvector_row} only concerns $B_{ij} \in S \times \paren{V\setminus S}$ we can set $B_{ij}=0,\forall \set{i,j} \in \paren{V \setminus S \times V \setminus S}$. Since equation \prettyref{eq:duality_dual_variables_condition} already forces us to set $B_{ij} \in S \times S$ to be $0$ we have,
\begin{align}\label{eq:B_matrix}
    B=B_{S \times S} + B_{S \times \paren{V \setminus S}} + B_{\paren{V \setminus S} \times S} + B_{\paren{V \setminus S} \times \paren{V \setminus S}} = B_{S \times \paren{V \setminus S}} + B_{\paren{V \setminus S} \times S} \mper
\end{align}
We will next show that under the assumption about existence of such $B_{ij}$'s, how we can proceed towards satisfying the optimality conditions for dual. 

\begin{fact} \label{fact:eigenvec_basic}
Given $T$ to be some set of orthonormal eigenvectors of a symmetric matrix $M$ labeled as 
\begin{align*}
    T= \set{\mathbf{u}^{\paren{1}},\mathbf{u}^{\paren{2}},\hdots,\mathbf{u}^{\paren{n}}} ,
\end{align*}
to show that $M \succeq 0$ it is sufficient to show that ${\mathbf{u}^T}M{\mathbf{u}} \geq 0, \forall \mathbf{u} \in T$.
\end{fact}

\begin{lemma}\label{lem:dual_matrix_psd}
For $n,k,d, p$ satisfying  $\lambda_{\min}\paren{A_{\paren{V \setminus S} \times \paren{V \setminus S}}-p\one_{V\setminus S}\one_{V \setminus S}^T} \geq - \tau $, $k \geq \paren{\ffrac{512}{\alpha p^{\ffrac{7}{2}}}}\sqrt{n \log n}$, $\gamma \geq \ffrac{1}{2}+\alpha$ and for choice of $\tau = \paren{\ffrac{d\paren{1/2-\alpha}}{\paren{1/2+\alpha}}}$ where
$p \geq 5\paren{\ffrac{\paren{\log k}}{\gamma^4 k}}^{\ffrac{1}{6}}$ and if there exists a $B$ satisfying equation \prettyref{eq:eigenvector_row} such that $\norm{B} \leq \ffrac{896\sqrt{n\paren{\log k}}}{p^{\ffrac{5}{2}}}$, with high probability (over the randomness of the input) the dual matrix $Y \succeq 0$.
Additionally, if the graph is connected, we have that $\lambda_2\paren{Y}>0$.
\end{lemma}

\begin{proof}
We will proceed using \prettyref{fact:eigenvec_basic} and show that  $\mathbf{u}^T Y \mathbf{u} \geq 0$, 
for all eigenvectors $\mathbf{u} \in T$ of the dual matrix $Y$
with high probability (over the randomness of the input).
We start with the vectors in the set $P_{-\tau}$ which are now also eigenvectors of $Y$ (after padding them with $0$'s). These are already orthonormal since $A_{S \times S}$ is symmetric. Also these eigenvectors have eigenvalue $d+ \lambda_l \geq 0$ and therefore,
\begin{align*}
  {\mathbf{v}^{\paren{l}}}^T Y {\mathbf{v}^{\paren{l}}} = \inprod{\mathbf{v}^{\paren{l}},Y \mathbf{v}^{\paren{l}}} = \paren{d+\lambda_l} \norm{\mathbf{v}^{\paren{l}}}^2 \geq 0,\quad  \forall \, \mathbf{v}^{\paren{l}} \in P_{-\tau} \mper
\end{align*}

Since the dual matrix $Y$ is symmetric, we can extend the set of vectors $P_{-\tau}$ to a complete orthonormal set of $n$ eigenvectors.
We include an eigenvector $\mathbf{x} \in T \setminus P_{-\tau}$ if $\inprod{\mathbf{x},\mathbf{v}^{(l)}}=0,\forall \mathbf{v}^{(l)} \in P_{-\tau}$. Since $\mathbf{v}^{(l)}_i=0$ for all $i \notin S$, this also implies that
$\inprod{\mathbf{x}_S,\mathbf{v}^{\paren{l}}}=0, \forall \, \mathbf{v}^{\paren{l}} \in P_{-\tau}$ and hence for such a vector $\mathbf{x} \in T \setminus P_{-\tau}$ we have,
\begin{align} \label{eq:qf_term1}
    \mathbf{x}^T A_{S \times S}\mathbf{x} = \mathbf{x}_{S}^T A_{S \times S} \mathbf{x}_S \geq -\tau\norm{\mathbf{x}_{S}}^2 \mper
\end{align}
for $\tau \geq 0$. Now we examine the quadratic form (w.r.t the dual matrix) for such vectors $\mathbf{x}$ in the subspace formed by vectors of the set $T \setminus P_{-\tau}$,
\begin{align*}
    \hspace{-1em}\mathbf{x}^T Y \mathbf{x} 
    &= \mathbf{x}^T\paren{A_{S \times S} + A_{V \setminus S \times V \setminus S} + p\paren{\one\one^T-\one_{S}\one_{S}^T - \one_{V\setminus S}\one_{V\setminus S}^T} + R + dI + B} \mathbf{x}\\
    &\geq \mathbf{x}_S^T\paren{A_{S \times S} -p\one_S\one_{S}^T}\mathbf{x}_S 
    + \mathbf{x}_{V\setminus S}^T\paren{A_{\paren{V\setminus S} \times \paren{V \setminus S}} -p{\one_{V\setminus S}\one_{V \setminus S}^T}}\mathbf{x}_{V \setminus S} \\
    &\qquad\qquad\qquad\qquad\qquad\qquad\qquad\qquad+ \paren{d-\norm{R}-\norm{B}}\norm{\mathbf{x}}^2\\
    &\geq \underbrace{\mathbf{x}_S^T\paren{A_{S \times S} -p\one_S\one_{S}^T}\mathbf{x}_S}_{=T_1}
    + \lambda_{\min}\underbrace{\paren{A_{\paren{V \setminus S} \times \paren{V \setminus S}}-p\one_{V \setminus S}\one_{V \setminus S}^T}}_{=T_2}\norm{\mathbf{x}_{{V \setminus S}}}^2\\
    \numberthis \label{eq:quadratic_form}
   &\qquad\qquad\qquad\qquad\qquad\qquad\qquad\qquad+ \underbrace{\paren{d-\norm{R}-\norm{B}}}_{=T_3} \norm{\mathbf{x}}^2 \mper
\end{align*}

We start by considering the term $T_1$. We already argue in equation \prettyref{eq:qf_term1} that the term $\mathbf{x}_S^TA_{S \times S}\mathbf{x}_S \geq -\tau$. The $-p\one_S\one_S^T$ term shifts only the $\one_S$ eigenvector of $A_{S \times S}$ and the corresponding eigenvalue is $d-pk$. Now as long as $d-pk \geq -\tau$, we have that the first term is at least $-\tau\norm{\mathbf{x}_S}^2$.

Next, we consider the term $T_2$. By our assumption on the smallest eigenvalue of $A_{\paren{V \setminus S} \times \paren{V \setminus S}}-p\one_{V \setminus S}\one_{V\setminus S}^T$ we have that,
\begin{align} \label{eq:quadratic_form_p1}
   \mathbf{x}_{V \setminus S}^T\paren{A_{\paren{V \setminus S} \times \paren{V \setminus S}}-p\one_{V \setminus S}\one_{V \setminus S}^T}\mathbf{x}_{V \setminus S} \nonumber
   &\geq \lambda_{\min}\paren{A_{\paren{V \setminus S} \times \paren{V \setminus S}}}\norm{\mathbf{x}_{V\setminus S}}^2\\
   &\geq -\tau\norm{\mathbf{x}_{V \setminus S}}^2 \mper
\end{align}
\noindent
Therefore using equations \prettyref{eq:qf_term1}, \prettyref{eq:quadratic_form} and \prettyref{eq:quadratic_form_p1} we obtain that,
\begin{align}\label{eq:qf_non_positive}
    \mathbf{x}^TY\mathbf{x} \geq \paren{d-\norm{R}-\norm{B}}\norm{\mathbf{x}}^2 -\tau\norm{\mathbf{x}_S}^2 - \tau\norm{\mathbf{x}_{V \setminus S}}^2= \paren{d-\tau-\norm{R}-\norm{B}}\norm{\mathbf{x}}^2 \mper
\end{align}

We recall that $\norm{R} \leq 2\sqrt{n}$ (\prettyref{claim:random_matrix_norm}) and for our choice of $\tau$ and for 
$B$ satisfying equation \prettyref{eq:eigenvector_row} for each $\mathbf{v}^{(l)} \in P_{-\tau}$ such that
$\norm{B} \leq \ffrac{896\sqrt{n(\log k)}}{p^{\ffrac{5}{2}}}$, in the regimes of $k,n,p$ as stated, we have that,
\begin{align*}
    d-\tau = d\paren{1-\dfrac{1/2-\alpha}{1/2+\alpha}} = \dfrac{2\alpha d}{1/2+\alpha} = \frac{2\alpha\gamma pk}{1/2+\alpha} \geq 2\alpha pk \geq \dfrac{1024}{p^{\ffrac{5}{2}}}\sqrt{n\log n} > \norm{B}+\norm{R} \mper
\end{align*}
Therefore, we obtain that $\mathbf{x}^TY\mathbf{x} \geq 0$ in equation \prettyref{eq:qf_non_positive} and the dual matrix $Y \succeq 0$.

Now we consider the scenario when the planted bipartite graph is connected.
For eigenvectors $\mathbf{v}^{(l)}\notin P_{-\tau}$, it follows from discussion above and equation \prettyref{eq:qf_non_positive} that the quadratic form is strictly positive.
Next, we consider the eigenvectors  $\mathbf{v}^{(l)} \in P_{-\tau}$ having eigenvalue $d+\lambda_l$.
If the planted bipartite graph is connected, it follows from basic spectral graph theory that the vector $\one_{S_1}-\one_{S_2}$ is the only eigenvector of $A_{S \times S}$ with eigenvalue $-d$.
Therefore, in our construction it is the only eigenvector of $Y$ with eigenvalue $0$.
Therefore we have that $\lambda_2(Y)>0$.
\end{proof}
\noindent
We next present the algorithm based on the guarantees provided by \prettyref{lem:dual_matrix_psd} as,

\RestyleAlgo{ruled}
\begin{minipage}{\linewidth}
\hspace{-1.75em}
	\begin{algorithm}[H]
		\caption{}
		\label{alg:two}
		\begin{algorithmic}[1]
			\REQUIRE $G=(V, E)$ sampled as per \prettyref{def:semi-random_model} ($d = \gamma pk, \gamma \geq \ffrac{2}{3}$).
			\ENSURE The set of vertices in planted bipartite graph $\mathcal{S}$ (with high probability).
			\STATE Solve \prettyref{sdp:primal}.
			\STATE Return $\mathcal{S} = \set{i : \norm{\mathbf{x}_i} > 0}$
		\end{algorithmic}
	\end{algorithm}
\end{minipage}

\subsection{Setting dual variables using Farkas' Lemma}
\label{sec:dual_feasible}
Now we are left with the task of showing that equation \prettyref{eq:eigenvector_row} indeed has a solution. Also, as discussed in the proof overview \prettyref{sec:proof_overview} (formally in \prettyref{sec:optimal_dual_construction}), we want the dual solution to satisfy the constraint $\norm{B} = \bigO_{p}\paren{\sqrt{n\log k}}$.  \prettyref{lem:dual_matrix_psd} shows that we are done provided there exists a choice of dual variables $B$ which satisfies this constraint. Turns out, this constraint is implied if $B_{ij} \leq t = \bigO_{p}\paren{\sqrt{\ffrac{\log k}{k}}}$ as shown below.

\begin{lemma}\label{lem:entrywise_bound_general}
For $0 \leq B_{ij} \leq t, \,\forall (i,j) \in \paren{S \times \paren{V \setminus S}} \cup \paren{\paren{V \setminus S} \times S}$ and $B_{ij}=0,\forall (i,j) \in \paren{S \times S} \cup \paren{\paren{V \setminus S} \times \paren{V \setminus S}}$  we have that $\norm{B}_2 \leq 2t\sqrt{k\paren{n-k}} \mper$ 
\end{lemma}
\begin{proof}
\begin{align*}
    \norm{B}_2 
    &\leq \norm{B_{S,\paren{V\setminus S}}}_2 + \norm{B_{\paren{V\setminus S},S}}_2 = 2 \norm{B_{S,\paren{V\setminus S}}}_2 \leq 2\norm{B_{S,V \setminus S}}_F \\ &=2\sqrt{\sum_{i \in S, j \in V \setminus S}B_{ij}^2}  \leq 2t\sqrt{k\paren{n-k}} \mper
\end{align*}
\end{proof}

\begin{corollary} \label{cor:entrywise_bound}
If each of the dual variable $B_{ij}, \forall (i,j) \in \paren{S \times \paren{V \setminus S}} \cup \paren{\paren{V \setminus S} \times S}$ satisfies $B_{ij} \leq \paren{56L_{-\tau}/p^{\ffrac{3}{2}}}\sqrt{\ffrac{\paren{\log k}}{\gamma k}}$  we have that $\norm{B}_2 \leq \paren{\ffrac{896}{p^{\ffrac{5}{2}}}}\sqrt{n\log k}$.
\end{corollary}
\begin{proof}
We already argue in discussion preceding equation \prettyref{eq:B_matrix}  that $B_{ij}=0,\forall (i,j) \in \paren{S \times S} \cup \paren{\paren{V \setminus S} \times \paren{ V \setminus S}}$.
Using \prettyref{lem:entrywise_bound_general} and setting $t=\paren{56L_{-\tau}/p^{\ffrac{3}{2}}}\sqrt{\ffrac{\paren{\log k}}{\gamma k}}$ we have,
\begin{align*}
    \norm{B}_2 \leq 2t\sqrt{k\paren{n-k}} \leq \dfrac{112L_{-\tau}}{p^{\ffrac{3}{2}}\sqrt{\gamma}}\sqrt{n\log k} \mper
\end{align*}
From \prettyref{fact:threshold_rank} we have that, $L_{-\tau} \leq  \ffrac{\gamma pk^2}{{2\tau}^2}$. Our choice of $\tau$ in \prettyref{lem:dual_matrix_psd} is such that $\tau \geq d/2$. This is because we chose $\tau = \ffrac{(0.5-\alpha)d}{(0.5 + \alpha)}$ where $\alpha \in (0,1/6]$. Hence,
\begin{align*}
    \norm{B}_2 &\leq \frac{112L_{-\tau}}{p^{3/2}\sqrt{\gamma}}\sqrt{n \log k} \leq \frac{112 \gamma pk^2}{2\tau^2p^{3/2}\sqrt{\gamma}}\sqrt{n \log k} \leq \frac{224\sqrt{\gamma}k^2}{\sqrt{p}d^2}\sqrt{n\log k}\\
    &= \frac{224\sqrt{\gamma}k^2}{\sqrt{p}(\gamma pk)^2}\sqrt{n \log k} = \frac{224}{p^{5/2}}.\frac{1}{\gamma^{3/2}}\sqrt{n \log k} \leq \frac{896}{p^{5/2}}\sqrt{n \log k}
\end{align*}
where the last inequality follows from the fact that $\gamma \geq 0.5 + \alpha \geq 0.5$.
\end{proof}

Next we aim to show that there exists a solution to $B_{ij}$'s which satisfies equation \prettyref{eq:eigenvector_row} and the criteria in \prettyref{cor:entrywise_bound} which eventually meets the hypothesis of \prettyref{lem:dual_matrix_psd}. Now, we consider the collection of linear systems $\{\cF_j\}_{j \in V \setminus S}$ (from our Proof Overview \prettyref{sec:proof_overview_pseudorandom}).

For the convenience of the reader, we now recall parts of our discussion from the proof overview. Recall $\cF_j$ is a linear system of the form $W_{-\tau}\mathbf{x}=\mathbf{b}$ where $\mathbf{b}$ is a row vector of size $L_{-\tau} \times 1$ and has entries given by
\begin{align} \label{eq:define_vector_b}
    b_l = - \sum_{i \in S}{A_{ij}v^{\paren{l}}_i}, \quad \forall l \in [L_{-\tau}]
\end{align}
and  $\mathbf{x}$ here is a row vector of size $k \times 1$ where the entry $x_i = B_{ij}$ (recall that we have fixed a $j \in V \setminus S$). However since $B_{ij}$'s are not arbitrary variables but dual variables of \prettyref{sdp:dual}, these are required to be non-negative and should only be defined for $i \in N(j)$. Thus, for any $j \in V \setminus S$, it is convenient to set $B_{ij} = 0$ for $i \not\in N(j)$. 

Now, let $\tilde W_{-\tau}$ denote the submatrix after removing the columns corresponding to $i \notin N(j)$ and $t$ to be the absolute bound on the entries of $B$ matrix (as desired in \prettyref{cor:entrywise_bound}). We thus consider the following feasibility LP formulation for this problem.

\begin{mybox}
		\begin{LP}
		\label{lp:primal}
		\begin{align}
			\label{eq:lp11}
			&\tilde W_{-\tau}\mathbf{x}=\mathbf{b}\\
			\label{eq:lp21}
			&0 \leq \mathbf{x} \leq t\one \mper
		\end{align}
		\end{LP}
\end{mybox}

The feasibility for such LPs is typically characterized by the Theorem of Alternatives (e.g., Farkas' Lemma). The standard variants for these deal with either the equality constraints or the inequality constraints. Here, our \prettyref{lp:primal} has mixed constraints, but we can derive a Farkas' Lemma style Theorem of Alternatives (along the lines of \cite{BV04}) as.

\begin{fact}[Folklore] \label{fact:farkas_variant}
For a fixed $\mathbf{u} \in \mathbb{R}^{L_{-\tau}}$, exactly one out of these two systems of linear equations is feasible,
\begin{enumerate}
    \item $\set{\mathbf{x}:C\mathbf{x}=\mathbf{f},\ \mathbf{0} \leq  \mathbf{x} \leq \mathbf{u}}$.
    \item $\set{\mathbf{y}:C^T\mathbf{y} + \mathbf{z}  \geq 0,\mathbf{f}^T\mathbf{y} + \mathbf{u}^T\mathbf{z}<0,\mathbf{y} \in \mathbb{R}^{L_{-\tau}},\mathbf{z} \geq 0} \mper$
\end{enumerate}
\end{fact}
\begin{proof}
For completeness, we give a proof along the lines of proof for the standard variants of Farkas' Lemma in \prettyref{app:farkas_variant_proof}.
\end{proof}

\begin{corollary} \label{cor:farkas_primal_feasible}
The primal \prettyref{lp:primal} is feasible iff 
\begin{align} \label{eq:farkas_condition}
    \forall \mathbf{y} \in \mathbb{R}^{L_{-\tau}},\forall \mathbf{z}\geq 0, \tilde W^T\mathbf{y} + \mathbf{z} \geq 0 \implies \mathbf{b}^T\mathbf{y}+t\inprod{\mathbf{z},\one} \geq 0 \mper
\end{align}
\end{corollary}
\begin{proof}
The above follows by setting $\mathbf{u}=t\one$, $\mathbf{f}=\mathbf{b}$ and $C=\tilde W$ in \prettyref{fact:farkas_variant}.
\end{proof}
 
We wish to compute a value of $t > 0$ such that the equation \prettyref{eq:farkas_condition} holds. Proving this seems to require a better understanding of the structure of eigenvectors in the set $P_{-\tau}$ (which we called as $\tau$-\textit{threshold rank eigenvectors}). Therefore, next (in \prettyref{lem:eigenvec_l_inf_norm} and \prettyref{lem:sum_spectral_embed_bound}) we prove some useful properties of these eigenvectors which will be used in the analysis. Throughout the rest of the section we will assume that the eigenvectors $\set{\mathbf{v}^{\paren{1}},\hdots,\mathbf{v}^{\paren{L_{-\tau}}}} \in P_{-\tau}$ have unit norm.

\begin{lemma}
\label{lem:eigenvec_l_inf_norm}
For an eigenvector $\mathbf{v}^{\paren{l}} \in P_{-\tau}$, where $\tau >d/2$ we have that, $\norm{\mathbf{v}^{\paren{l}}}_{\infty} \leq \dfrac{2}{\sqrt{d}}$.
\end{lemma}

\begin{proof}
Since $\mathbf{v}^{\paren{l}}$ is an eigenvector of $\restrict{A}{S \times S}$ we have that,
\[ \restrict{A}{S \times S}\mathbf{v}^{\paren{l}} = \lambda_l\mathbf{v}^{\paren{l}} \mper \label{eq:eigenvector_equation} \numberthis\]
We compare the vectors in equation \prettyref{eq:eigenvector_equation} component wise and we have that,
\begin{align}
    \sum_{i \in N(j)}{\mathbf{v}^{\paren{l}}_i}= \lambda_l\mathbf{v}^{\paren{l}}_j \quad \forall j \in S \mper \label{eq:componentwise_eigenvectors} 
\end{align} 
Take absolute value on both sides and use Cauchy–Schwarz. This gives
\begin{align*} \abs{\lambda_l\mathbf{v}^{\paren{l}}_j} = \abs{\sum_{i \in N(j)}{\mathbf{v}^{\paren{l}}_i}} = \abs{\inprod{\one_{N(j)},\mathbf{v}^{\paren{l}}}} \leq \sqrt{d}\sqrt{\norm{\mathbf{v}^{\paren{l}}}^2} \leq \sqrt{d} \mper
\end{align*}
This allows us to give an upper bound on the $l_{\infty}$ norm of the eigenvector $\mathbf{v}^{\paren{l}}$ by comparing the left and right hand side as,
\begin{align}
    \abs{{v}^{\paren{l}}_j} \leq \dfrac{\sqrt{d}}{\abs{\lambda_l}} \leq \dfrac{2}{\sqrt{d}} \mper \label{eq:l_inf_norm_bound}
\end{align} 
The last inequality holds since for $\tau \geq d/2$, we have $\lambda_l \in [-d/2,-d]$ and hence $\abs{\lambda_l} \geq d/2$.
\end{proof}

\begin{lemma} \label{lem:sum_spectral_embed_bound}
For a fixed $j \in V \setminus S$ and for $\set{\mathbf{w}^{\paren{i}}\vert i \in N(j)}$, we have that with probability at least $1-\bigO\paren{\ffrac{1}{k^8}}$,
\begin{align*}
    \norm{\sum_{i \in N(j)}{\mathbf{w}^{\paren{i}}}}_2 \leq 3\sqrt{{L_{-\tau}\log k}} \mper
\end{align*}
\end{lemma}

\begin{proof}
To show the claim above, we first show that $\norm{\sum_{i \in N(j)}{\mathbf{w}^{\paren{i}}}}_{\infty} \leq 3\sqrt{{\log k}}$. Now since $\set{\mathbf{v}^{\paren{1}},\hdots,\mathbf{v}^{\paren{L_{-\tau}}}}$ were orthogonal to $\one_S$ vector to start with, $\sum_{i \in S}\mathbf{w}^{\paren{i}}=0$. For an arbitrary $r\in [L_{-\tau}]$ we bound $\abs{\sum_{i \in N(j)}{{w}^{\paren{i}}_r}}$.
Since $\sum_{i \in S}{w}^{\paren{i}}_r=0$ we have that,
\[ \E\left[\sum_{i \in N(j)}{{w}^{\paren{i}}_r}\right] = 0 \mper\]
Then we can use Hoeffding bounds (\prettyref{fact:hoeffding_bound}) to bound as,
\begin{align*}
    \Pr{\abs{\sum_{i \in N(j)}{{w}^{\paren{i}}_r}} \geq 3\sqrt{\log k}} 
    &\leq 2\exp\paren{-\dfrac{9\log k}{\sum_{i=1}^k{{w^{\paren{i}}_r}^2}}} = 2\exp\paren{-\dfrac{9\log k}{\sum_{i=1}^k{{v^{\paren{r}}_i}^2}}}\\
    &= 2\exp\paren{-9{\log k}} = \dfrac{2}{k^9}\mper
\end{align*}
For any $r \in [L_{-\tau}]$, we can use a union bound over $r \in [L_{-\tau}]$ to claim that,
\begin{align*}
    \Pr{\abs{\sum_{i \in N(j)}{{w}^{\paren{i}}_r}} \geq 3\sqrt{\log k}} \leq \frac{2L_{-\tau}}{k^9} \leq \frac{2k}{k^9}=\frac{2}{k^8}
\end{align*}
Therefore with probability $\geq 1-\bigO\paren{\ffrac{1}{k^8}}$, we have that $\norm{\sum_{i \in N(j)}{\mathbf{w}^{\paren{i}}}}_2 \leq 3\sqrt{L_{-\tau}{\log k}} \mper$
\end{proof}

\begin{proposition} \label{prop:farkas_apply}
For choice of $t = \paren{\ffrac{56L_{-\tau}}{p^{\ffrac{3}{2}}}}\paren{\sqrt{\ffrac{\paren{\log k}}{\gamma k}}} $ and for regimes of $p \geq 5\paren{\ffrac{ \paren{\log k}}{\gamma^4 k}}^{\ffrac{1}{6}}$, with high probability (over the randomness of the input instance),
\begin{align*}
    \forall \, \mathbf{y} \in \mathbb{R}^{L_{-\tau}},\forall \, \mathbf{z}\geq 0, \tilde W_{-\tau}^T\mathbf{y} + \mathbf{z} \geq 0 \implies \mathbf{b}^T\mathbf{y}+t\inprod{\mathbf{z},\one} \geq 0 \mper
\end{align*}
\end{proposition}
\begin{proof}
For the sake of contradiction, suppose there exists a $\mathbf{y} \in \mathbb{R}^{L_{-\tau}}$ and $\mathbf{z} \geq 0$ such that for our choice of $t$, $\tilde W_{-\tau}^T\mathbf{y} + \mathbf{z} \geq 0$ and $\mathbf{b}^T\mathbf{y}+t\inprod{\mathbf{z},\one} < 0$.
Now, we choose $\mathbf{z}'$ by setting $z'_i=\max\set{0,-\inprod{\mathbf{w}^{(i)},\mathbf{y}}}$ and for the same value of $t$ and $\mathbf{y}$, the assumptions for contradiction (i.e. $\tilde W_{-\tau}^T\mathbf{y} + \mathbf{z}' \geq 0$ and $\mathbf{b}^T\mathbf{y}+t\inprod{\mathbf{z}',\one} < 0$) continue to hold. This is because we can write the condition in the contradiction 
$ \tilde W_{-\tau}^T\mathbf{y}+\textbf{z} \geq 0 \text{ in the form } \paren{\tilde W_{-\tau}^T\mathbf{y}}_i +z_i \geq 0
     \text{ and}
      \text{ we have } z_i \geq -\inprod{\mathbf{w}^{\paren{i}},\mathbf{y}}$.
Therefore, by construction, we have $z'_i = \max\set{0,-\inprod{\mathbf{w}^{i},\mathbf{y}}} \leq z_i$ entrywise. Hence, we have $0 \geq \tilde W^T_{-\tau}\mathbf{y}+\mathbf{z} \geq \tilde W^T_{-\tau}\mathbf{y}+\mathbf{z}' $. Also, since $t>0$ we get,
\begin{align*}
    \mathbf{b}^T\mathbf{y}+t\inprod{\mathbf{z}',\one} = \mathbf{b}^T\mathbf{y} + t\paren{\sum_i z^{'}_i} \leq \mathbf{b}^T\mathbf{y} + t\paren{\sum_i z_i} = \mathbf{b}^T\mathbf{y} + t\inprod{\mathbf{z},\one} <0 \mper
\end{align*}

\noindent
Now, we will show a contradiction for this value of $t,\mathbf{y}$ and $\mathbf{z}'$. 
We consider the term $\mathbf{b}^T\mathbf{y}$ and using equation \prettyref{eq:define_vector_b}, we express it as,
\begin{align*} \label{eq:spectral_embedding} \numberthis
\mathbf{b}^T\mathbf{y}
&= \sum_{r \in [L_{-\tau}]}{b_ry_r} = -\sum_{r \in [L_{-\tau}]}\sum_{i \in N(j)}v^{\paren{r}}_i y_r = -\sum_{i \in N(j)}\sum_{r \in [L_{-\tau}]}v^{\paren{r}}_i y_r\\
&= -\sum_{i \in N(j)}\sum_{r \in [L_{-\tau}]}w^{\paren{i}}_r y_r = -\sum_{i \in N(j)}\inprod{\mathbf{w}^{\paren{i}},\mathbf{y}} \mper
\end{align*}
Using the value of $\mathbf{z}'$ as above i.e  $z^{'}_i=\max\set{0,-\inprod{\mathbf{w}^{\paren{i}},\mathbf{y}}}$, we can rewrite the condition $\mathbf{b}^T\mathbf{y}+t\inprod{\mathbf{z}',\one}<0$ as,
\begin{align} \label{eq:farkas_no_solution}
    \sum_{i \in N(j)}\inprod{\mathbf{w}^{\paren{i}},\mathbf{y}} + t\sum_{i\in N(j)}{\min\set{0,{\inprod{\mathbf{w}^{\paren{i}},\mathbf{y}}}}} >0 \mper
\end{align}
To finish the proof by contradiction we will show that equation \prettyref{eq:farkas_no_solution} does not hold with high probability (over the randomness of the input) for our choice of $t$.
Without loss of generality we assume $\norm{\mathbf{y}}=1$ and proceed to bound the first term in the expression on left hand side of equation \prettyref{eq:farkas_no_solution} as,
\begin{align} \label{eq:local_first_term_bound}
    \sum_{i \in N(j)}\inprod{\mathbf{w}^{\paren{i}},\mathbf{y}} = \inprod{\mathbf{y},\sum_{i \in N(j)}{\mathbf{w}^{\paren{i}}}} \leq \norm{\mathbf{y}}\norm{\sum_{i \in N(j)}{\mathbf{w}^{\paren{i}}}} \leq 3\sqrt{{L_{-\tau}\log k}}
\end{align}
where the last inequality follows from \prettyref{lem:sum_spectral_embed_bound}.
In \prettyref{lem:farkas_condition_second_term} we give an upper bound on the second term as
\begin{align} \label{eq:local_second_term_bound}
    \sum_{i\in N(j)}{\min\set{0,{\inprod{\mathbf{w}^{\paren{i}},\mathbf{y}}}}} \leq -\paren{\ffrac{{p}}{16}}\sqrt{\ffrac{d}{L_{-\tau}}} \mper
\end{align}

We do a union bounds on the bounds obtained from \prettyref{lem:sum_spectral_embed_bound} and \prettyref{lem:farkas_condition_second_term} (which hold with probability $\geq 1-\bigO\paren{1/k^4} \geq 1-\bigO\paren{1/n^2}$) and we get that for all $j \in V \setminus S$, the bounds in \prettyref{lem:sum_spectral_embed_bound} and \prettyref{lem:farkas_condition_second_term} hold with high probability (over the randomness of the input). Since the bounds in equations \prettyref{eq:local_first_term_bound} and \prettyref{eq:local_second_term_bound} come from \prettyref{lem:sum_spectral_embed_bound} and \prettyref{lem:farkas_condition_second_term}, we also have them hold for all $j \in V \setminus S$ with high probability (over the randomness of the input instance).
Then for any $j \in V \setminus S$ and our choice of $t = \paren{\ffrac{56L_{-\tau}}{p^{\ffrac{3}{2}}}}\paren{\sqrt{\ffrac{\paren{\log k}}{\gamma k}}}$ in equation \prettyref{eq:farkas_no_solution} we get,
\begin{align*}
    &3\sqrt{{L_{-\tau}\log k}} - \paren{\dfrac{56L_{-\tau}\sqrt{\log k}}{p^{\ffrac{3}{2}}\sqrt{\gamma k}}}\paren{\dfrac{{p}\sqrt{d}}{16\sqrt{L_{-\tau}}}} \text{ and using $d=\gamma pk$ we simplify and obtain}\\
    &3\sqrt{{L_{-\tau}\log k}} - \paren{\dfrac{56L_{-\tau}\sqrt{\log k}}{p^{\ffrac{3}{2}}\sqrt{\gamma k}}}\paren{\dfrac{{p^{\ffrac{3}{2}}}\sqrt{\gamma k}}{16\sqrt{L_{-\tau}}}} <0 \text{ and hence equation \prettyref{eq:farkas_no_solution} doesn't hold.}
\end{align*}
\end{proof}

Towards upper bounding the second term (as in \prettyref{lem:farkas_condition_second_term}), we begin with a crucial observation in \prettyref{lem:matrix_concentration_bound}.

\begin{lemma} \label{lem:matrix_concentration_bound}
For an arbitrary unit vector $\mathbf{y} \in \mathbb{R}^{L_{-\tau}}$ and a fixed $j \in V \setminus S$, for $p \geq 5\paren{\ffrac{\log k}{\gamma^4 k}}^{\ffrac{1}{6}}$, we have that with probability $\geq 1-\bigO\paren{\ffrac{1}{k^4}}$,
\begin{align}\label{eq:projection_rewrite}
    \sum_{i \in N(j)}{\inprod{\mathbf{y},\mathbf{w}^{\paren{i}}}^2} \geq \dfrac{p}{2} \mper
\end{align}
\end{lemma}
\begin{proof}
\begin{align}\label{eq:relate_isotropicity}
    \sum_{i \in N(j)}{\inprod{\mathbf{y},\mathbf{w}^{\paren{i}}}^2} = \sum_{i \in N(j)}{\paren{\mathbf{y}^T\mathbf{w}^{\paren{i}}}}\paren{{\mathbf{w}^{\paren{i}}}^T\mathbf{y}} = \mathbf{y}^T\paren{\sum_{i \in N(j)}{\mathbf{w}^{\paren{i}}{\mathbf{w}^{\paren{i}}}^T}}\mathbf{y} \mper 
\end{align}
\begin{fact} \label{fact:matrix_expt}
For a fixed $j \in V \setminus S$, we let $M=\sum_{i \in N(j)}{\mathbf{w}^{\paren{i}}{\mathbf{w}^{\paren{i}}}^T}$, a matrix of size $L_{-\tau} \times L_{-\tau}$ then,
\begin{align*}
    \E\Brac{M_{rs}}=0 \text{ for } r\neq s \text{ and } \E\Brac{M_{rr}}=p \mper
\end{align*}
\end{fact}

\begin{proof}
A proof of this can be found in \cite{LRTV11}. For completeness we give a proof in \prettyref{app:matrix_expt}
\end{proof}

\noindent
Therefore we conclude that $\E[M]=pI$.
Next, we show the concentration of the matrix $M$ by using Matrix Bernstein inequality, Theorem 6.1 in the work \cite{Tro12} restated here as,
\begin{fact}[Matrix Bernstein inequality]
For a sequence of independent,symmetric, and random matrices $\set{X_i\in \mathbb{R}^{k \times k}}_{i=1}^{k}$ where,
\begin{itemize}
    \item $\E\left[X_i\right]=0$
    \item $\norm{X_i} \leq \rho$
    \item $\nu = \norm{\E\left[\sum_{i}\paren{X_i^2}\right]}$
    \end{itemize}   
we have that for all $\varepsilon \geq 0$,
    \[\Pr{\norm{\sum_{i}{X_i}} \geq \e } \leq 2k\exp\paren{\dfrac{-\e^2}{2\nu + 2\rho\e/3}} \mper\]
\end{fact}
\noindent
In our setting we let,
\begin{equation*}
Y_i=
    \begin{cases} 
	\mathbf{w}^{\paren{i}}{\mathbf{w}^{\paren{i}}}^T & \text{ if } i \in N(j) \\
	0 & \text{ otherwise }\mper
	\end{cases} 
\end{equation*}
and define the random matrices $X_i = Y_i - \E\left[Y_i\right]$ so that $\E\left[X_i\right]=0$. Writing this explicitly we have,
\begin{equation*}
X_i=
    \begin{cases} 
	(1-p)\mathbf{w}^{\paren{i}}{\mathbf{w}^{\paren{i}}}^T & \text{ if } i \in N(j)\\
	-p\mathbf{w}^{\paren{i}}{\mathbf{w}^{\paren{i}}}^T & \text{ otherwise }\mper
	\end{cases} 
\end{equation*}
From above we can see that,
\begin{align*}
    \norm{X_i} 
    \leq \max\set{\norm{p\mathbf{w}^{\paren{i}}{\mathbf{w}^{\paren{i}}}^T},\norm{(1-p)\mathbf{w}^{\paren{i}}{\mathbf{w}^{\paren{i}}}^T}}  \leq \norm{\mathbf{w}^{\paren{i}}}^2 
    \leq \dfrac{2L_{-\tau}}{d} 
\end{align*}
where we have used the fact that $\mathbf{w}^{\paren{i}}\mathbf{w}^{{\paren{i}}^T}$ is a rank one matrix and hence $\norm{\mathbf{w}^{\paren{i}}\mathbf{w}^{{\paren{i}}^T}} \leq \norm{\mathbf{w}^{\paren{i}}}^2$,
and the last inequality holds since $\norm{\mathbf{w}^{(i)}}_{\infty} \leq 2/\sqrt{d}$ (using \prettyref{lem:eigenvec_l_inf_norm}).
Hence we choose value of $\rho=\ffrac{2L_{-\tau}}{d}$. Next we bound the variance $\nu$ as,
\begin{equation} \label{eq:bound_matrix_variance}
    \nu = \norm{\E\left[\sum_{i\in S}{X_i^2}\right]} = \norm{\sum_{i \in S}{\E\left[X_i^2\right]}} \leq \sum_{i \in S}\norm{\E\left[X_i^2\right]} \mper
\end{equation}
To compute $\E\left[X_i^2\right]$ we note that,
\begin{equation*}
   X_i^2 =  \begin{cases} 
	(1-p)^2\mathbf{w}^{\paren{i}}{\mathbf{w}^{\paren{i}}}^T\mathbf{w}^{\paren{i}}{\mathbf{w}^{\paren{i}}}^T & \text{ if } i \in N(j) \\
	p^2\mathbf{w}^{\paren{i}}{\mathbf{w}^{\paren{i}}}^T\mathbf{w}^{\paren{i}}{\mathbf{w}^{\paren{i}}}^T & \text{ otherwise } \mper
	\end{cases}
\end{equation*}
Therefore $\E\left[X_i^2\right]= p(1-p)\mathbf{w}^{\paren{i}}{\mathbf{w}^{\paren{i}}}^T\mathbf{w}^{\paren{i}}{\mathbf{w}^{\paren{i}}}^T$ and using this in equation \prettyref{eq:bound_matrix_variance} we get,
\begin{align} \label{eq:nu_calculations}
    \nu  &\leq \sum_{i \in S}{\norm{\E\Brac{X_i^2}}} \leq p(1-p)\sum_{i \in S}\norm{\mathbf{w}^{\paren{i}}\mathbf{w}^{{\paren{i}}^T}}\norm{\mathbf{w}^{\paren{i}}\mathbf{w}^{{\paren{i}}^T}}\\
    &\leq \sum_{i \in S}\norm{\mathbf{w}^{\paren{i}}}^4 = k\norm{\mathbf{w}^{\paren{i}}}^4 \leq \dfrac{16{L_{-\tau}}^2k}{d^2} \mper
\end{align}

Using the value of $\nu,\rho$ and the fact that $\E\Brac{\sum_{i \in N(j)}{\mathbf{w}^{\paren{i}}\mathbf{w}^{{\paren{i}}^T}}}=pI$ and choosing $\e = p/2$ we have that,
\begin{align*} \label{eq:matrix_concentration_bound}
    \Pr{\norm{\sum_{i \in N(j)}{\mathbf{w}^{\paren{i}}}{\mathbf{w}^{\paren{i}}}^T -pI}  \geq \dfrac{p}{2}} 
    &\leq  2k\exp\paren{-\dfrac{\paren{\ffrac{p^2}{4}}}{2\paren{\ffrac{16{L_{-\tau}}^2k}{d^2}}+\paren{\ffrac{2pL_{-\tau}}{3d}}}}\\
    &\leq 2k\exp\paren{-\dfrac{p^2/4}{2\paren{32L^2_{-\tau}k}/d^2}}  \\
    &\leq  2k\exp\paren{-\dfrac{\gamma^4 p^6 k}{1024}} \numberthis
\end{align*}
where the second inequality follows because
$\paren{\frac{32L_{-\tau}^2k}{d^2}} \geq \frac{2pL_{-\tau}}{3d}$ and the last inequality follows by using $\tau \geq d/2$ and corresponding $L_{-\tau}$ from \prettyref{fact:threshold_rank}.
For our parameter regimes, $\gamma \geq 1/2+\alpha \geq \ffrac{1}{2}$ and $p \geq 5\paren{\ffrac{\log k}{\gamma^4 k}}^{\ffrac{1}{6}}$, 
plugging equation \prettyref{eq:matrix_concentration_bound} in equation \prettyref{eq:projection_rewrite} and using Weyl's inequality (\prettyref{fact:weyl_inequality}), for any unit vector $\mathbf{y} \in \bbR^{L_{-\tau}}$ with probability at least $1-1/\bigO(1/k^4)$ we have,
\begin{align*}
    \sum_{i \in N(j)}\inprod{\mathbf{y},\mathbf{w}^{(i)}}^2 &= \mathbf{y}^T\paren{\sum_{i \in N(j)}\mathbf{w}^{(i)}{\mathbf{w}^{(i)}}^T}\mathbf{y} \geq \lambda_1\paren{\sum_{i \in N(j)}\mathbf{w}^{(i)}{\mathbf{w}^{(i)}}^T}\\
    &= \lambda_1\paren{\sum_{i \in N(j)}\mathbf{w}^{(i)}{\mathbf{w}^{(i)}}^T -pI + {pI}} \geq \lambda_{1}\paren{{pI}} + \lambda_1\paren{\sum_{i \in N(j)}\mathbf{w}^{(i)}{\mathbf{w}^{(i)}}^T-{pI}}\\
    &=p + \lambda_1\paren{\sum_{i \in N(j)}\mathbf{w}^{(i)}{\mathbf{w}^{(i)}}^T-\frac{pI}{2}} \geq p - \norm{\sum_{i \in N(j)}\mathbf{w}^{(i)}{\mathbf{w}^{(i)}}^T-\frac{pI}{2}} \geq \frac{p}{2} \mper
\end{align*}
Therefore, we have that 
\prettyref{lem:matrix_concentration_bound} holds with probability $\geq 1-\bigO\paren{\ffrac{1}{k^4}}$.
\end{proof}

With these bounds at hand we proceed to bound the second term in equation \prettyref{eq:farkas_no_solution} as,

\begin{lemma} \label{lem:farkas_condition_second_term}
For a unit vector $\mathbf{y} \in \mathbb{R}^{L_{-\tau}}$ and $j \in V \setminus S$ and for $p \geq 5\paren{\ffrac{ \paren{\log k}}{\gamma^4 k}}^{\ffrac{1}{6}}$ with probability $\geq 1-\bigO\paren{\ffrac{1}{k^4}}$ we have that,
\begin{align*}
    \sum_{i\in N(j)}{\min\set{0,{\inprod{\mathbf{w}^{\paren{i}},\mathbf{y}}}}} \leq -\dfrac{p}{16}\sqrt{\dfrac{d}{L_{-\tau}}} \mper
\end{align*}
\end{lemma}
\begin{proof}
We start by defining these two sets,
\begin{equation*}
    \mathcal{P}=\set{i \in N(j) : \inprod{\mathbf{y},\mathbf{w}^{\paren{i}}} \geq 0} 
\end{equation*}
\begin{equation*}
    \mathcal{N}=\set{i \in N(j) : \inprod{\mathbf{y},\mathbf{w}^{\paren{i}}} < 0} 
\end{equation*}
Clearly the summation over terms in $\mathcal{P}$ is $0$ so we focus on the terms in $\mathcal{N}$. We first consider the case when $\sum_{i \in \mathcal{N}}{\inprod{\mathbf{y},\mathbf{w}^{\paren{i}}}^2} \geq \ffrac{p}{4}$ and since $\abs{\inprod{\mathbf{y},\mathbf{w}^{\paren{i}}}} \leq \norm{\mathbf{y}}\norm{\mathbf{w}^{\paren{i}}} \leq \ffrac{2\sqrt{L{-\tau}}}{\sqrt{d}}$ (using \prettyref{lem:eigenvec_l_inf_norm}) we have that,
\begin{align*} 
\dfrac{p}{4} \leq \sum_{i \in \mathcal{N}}{\inprod{\mathbf{y},\mathbf{w}^{\paren{i}}}}^2 = \sum_{i \in \mathcal{N}}\abs{\inprod{\mathbf{y},\mathbf{w}^{\paren{i}}}}^2  \leq 2\sqrt{\dfrac{L_{-\tau}}{d}} 
\sum_{i \in \mathcal{N}}\abs{\inprod{\mathbf{y},\mathbf{w}^{\paren{i}}}} \mper
\end{align*}
and therefore $\sum_{i \in \mathcal{N}}{\inprod{\mathbf{y},\mathbf{w}^{\paren{i}}}} \leq -\paren{p/8}\sqrt{\ffrac{d}{L_{-\tau}}}$. 

Next we consider the case when $\sum_{i \in \mathcal{N}}{\inprod{\mathbf{y},\mathbf{w}^{\paren{i}}}^2} < \ffrac{p}{4}$. Now in this case, from \prettyref{lem:sum_spectral_embed_bound} we know that, $\norm{\sum_{i \in N(j)}{\mathbf{w}^{\paren{i}}}}_2 \leq 3\sqrt{{L_{-\tau}\log k}}$, we can write,
\begin{align} \label{eq:relate_bounds}
    \abs{\sum_{i \in N(j)}{\inprod{\mathbf{y},\mathbf{w}^{\paren{i}}}}} = \abs{{\inprod{\mathbf{y},\sum_{i \in N(j)}\mathbf{w}^{\paren{i}}}}}\leq \norm{\mathbf{y}}\norm{\sum_{i \in N(j)}{\mathbf{w}^{\paren{i}}}} \leq 3\sqrt{{L_{-\tau}\log k}} \mper
\end{align}
From equation \prettyref{eq:relate_bounds} we obtain that,
\begin{align}\label{eq:relate_bounds_properly}
     \sum_{i \in \mathcal{P}}{\inprod{\mathbf{y},\mathbf{w}^{\paren{i}}}} \leq - \sum_{i \in \mathcal{N}}{\inprod{\mathbf{y},\mathbf{w}^{\paren{i}}}} + 3\sqrt{{L_{-\tau}\log k}} \mper
\end{align}
Using equation \prettyref{eq:relate_bounds_properly} and $\sum_{i \in \mathcal{N}}{\inprod{\mathbf{y},\mathbf{w}^{\paren{i}}}^2} < \ffrac{p}{4}$ we have that,
\begin{align*} \label{eq:bound_second_term}
    \dfrac{p}{4} 
    &\leq \sum_{i \in \mathcal{P}}{\inprod{\mathbf{y},\mathbf{w}^{\paren{i}}}^2} \leq \norm{\mathbf{y}}\max_{i \in \mathcal{P}}\norm{\mathbf{w}^{\paren{i}}}\sum_{i \in \mathcal{P}}{\inprod{\mathbf{y},\mathbf{w}^{\paren{i}}}} \leq 2\sqrt{\dfrac{{L_{-\tau}}}{d}}\sum_{i \in \mathcal{P}}{\inprod{\mathbf{y},\mathbf{w}^{\paren{i}}}}\\
    &\leq - 2\sqrt{\dfrac{{L_{-\tau}}}{d}}\sum_{i \in \mathcal{N}}{\inprod{\mathbf{y},\mathbf{w}^{\paren{i}}}} + 6\sqrt{\dfrac{{L_{-\tau}}^2\log k}{d}} \numberthis \mper
\end{align*}
We wish that the term $6\sqrt{\ffrac{{L_{-\tau}}^2\paren{\log k}}{d}} \leq \ffrac{p}{8}$, since solving for $ \sum_{i \in \mathcal{N}}{\inprod{\mathbf{y},\mathbf{w}^{\paren{i}}}}$ in equation \prettyref{eq:bound_second_term} then gives us that,
\begin{align*}
    \sum_{i \in \mathcal{N}}{\inprod{\mathbf{y},\mathbf{w}^{\paren{i}}}} \leq -\dfrac{p}{16}\sqrt{\dfrac{d}{L_{-\tau}}}  \text{ and we are done.}
\end{align*}
\noindent
To show that the term $6\sqrt{\ffrac{{L_{-\tau}}^2\paren{\log k}}{d}} \leq \ffrac{p}{8}$, we start with a lower bound on $p$ as $p \geq 6\paren{\ffrac{\log k}{\gamma^3k}}^{1/5}$ and take powers on both side to get $p^5 \geq \frac{7776 \log k}{\gamma^3 k}$ which implies,
\begin{align*}
    p^2 &\geq \paren{\frac{7776}{\gamma^3p^3}}\frac{\log k}{k} =1944\paren{\frac{\gamma pk^2}{2(\gamma pk/2)^2}}^2 \frac{\log k}{\gamma pk} = 1944 \paren{\frac{\gamma pk^2}{2(d/2)^2}}^2\frac{\log k}{d} \paren{\text{Use } d=\gamma pk}\\
    \numberthis\label{eq:calculatint_p}
    &\geq 1944\paren{\frac{\gamma pk^2}{2\tau^2}}^2\frac{\log k}{d} \geq 1944L_{-\tau}^2\frac{\log k}{d} > \frac{384L_{-\tau}^2\log k}{d} \paren{\text{Use $\tau \geq \frac{d}{2}$ and \prettyref{fact:threshold_rank}}.}
\end{align*}
We note that the lower bound we assumed on $p$ is subsumed by the bound from \prettyref{lem:matrix_concentration_bound}.
Taking square root in equation \prettyref{eq:calculatint_p} we indeed get that $6\sqrt{\ffrac{{L_{-\tau}}^2\paren{\log k}}{d}} \leq \ffrac{p}{8}$.
\end{proof}

\subsection{Action of Adversary} \label{sec:adversary_action}
We recall that in \prettyref{step:adversary} of our model construction (\prettyref{def:semi-random_model}), we allow a monotone adversary to arbitrarily add edges between two vertices of $\paren{V \setminus S} \times \paren{V \setminus S}$. Here, we argue that despite the action of an adversary, our \prettyref{alg:two} still returns the planted set $S$. To show this, we use the argument from the work \cite{FK01} to recover the planted set $S$ under the action of adversary. 

\begin{lemma} \label{lem:handle_adversary}
Let $G$ be the graph obtained at end of \prettyref{step:rest_of_graph} and $\tilde G$ be the graph obtained after action of adversary on $G$ as per \prettyref{step:adversary}. Let $\OPT$ denote the cost of optimal solution, then we have $\OPT\paren{\tilde G} = \OPT\paren{G}$ and the solution in equation \prettyref{eq:integral} is the unique optimal solution.
\end{lemma}
\begin{proof}
We start by writing out an exact formulation of the problem as an integer quadratic program,
\begin{mybox}
        \begin{QP}
		\label{qp:original_formulation}
\[ \min \sum_{\set{i,j} \in E} x_ix_j\]
		\subjectto
		\begin{align}
			\label{eq:qp1}
			&\sum_{i \in V}x_i^2 = 1\\
			\label{eq:qp2}
			&x_ix_j\leq 0 &\forall \set{i,j} \in E\\
			&x_i^2 \in \set{0,1/k}
			\mper
		\end{align}
		\end{QP}
\end{mybox}

\noindent
We denote by $\mathsf{QPOPT}\paren{G}$ to be the minimum objective value to \prettyref{qp:original_formulation}. Now since this is an exact formulation for our problem we have that, $\OPT\paren{G} = \mathsf{QPOPT}\paren{G}$. Note that \prettyref{sdp:primal} is a relaxation to \prettyref{qp:original_formulation} and hence, \begin{align} \label{eq:relate_optimals}
\SDPOPT\paren{G} \leq \OPT\paren{G} = \mathsf{QPOPT}\paren{G} \mper
\end{align}

We will prove our claim by induction on the number of edges added by adversary in \prettyref{step:adversary}. 
We consider the base case first and let $G'$ to be the graph obtained after adding an edge to $G$. Since the integral solution \prettyref{eq:integral} is still a feasible solution to \prettyref{qp:original_formulation} for $G'$, using equality in equation \prettyref{eq:relate_optimals} we have that,
\begin{align*}
    \OPT\paren{G'} = \mathsf{QPOPT}\paren{G'} \leq -d \mper
\end{align*}
Also using inequality in equation \prettyref{eq:relate_optimals} we have that,
\begin{align*}
    -d-\frac{1}{k} \leq \SDPOPT\paren{G'} \leq \OPT\paren{G'},
\end{align*}
where first inequality follows from the fact that a solution of value strictly smaller than $-d-\frac{1}{k}$ to $\SDPOPT\paren{G'}$ would imply a solution of value strictly smaller that $-d$ to $\SDPOPT\paren{G}$. This is because if remove back the added edge from $G'$, the objective value falls by at most $1/k$ (due to the SDP constraint \prettyref{eq:sdp3}). Therefore we have that,
\begin{align*}
    -d-\frac{1}{k} \leq \OPT\paren{G'} \leq -d \mper
\end{align*}
Now since $\OPT\paren{G'}$ only takes values in steps of $1/k$, it can either be $-d$ or $-d-\frac{1}{k}$. However if $\OPT\paren{G'}=-d-\frac{1}{k}$, then $\SDPOPT\paren{G'} \leq -d-\frac{1}{k}$ and once we remove the edge back from $G'$ to obtain $G$, the SDP solution is still a feasible solution to $G$ with value less than or equal to $-d$. However this solution is different from the integral solution $\mathbf{g}\mathbf{g}^T$ since for the solution $\mathbf{g}\mathbf{g}^T$, $\SDPOPT\paren{G'}$ would have been $-d$ as well. Therefore, we have obtained a new solution to our \prettyref{sdp:primal}. However as argued in \prettyref{fact:optimality_conditions}, our SDP can only have a unique solution. Therefore we have that $\OPT\paren{G'}=-d=\OPT\paren{G}$.

\noindent
This contradiction above also proves that the SDP solution to $G'$ has to be $\mathbf{g}\mathbf{g}^T$ since otherwise we again get a new solution to \prettyref{sdp:primal} and a contradiction to \prettyref{fact:optimality_conditions} for $G$.

\noindent
Now $\tilde G$ is obtained by a sequence of such operations on $G$ and the argument above holds for each such operation, and therefore,
\begin{align*}
    \OPT\paren{G} = \OPT\paren{G'} = \hdots = \OPT\paren{\tilde G} \mper
\end{align*}
\end{proof}

We have now established everything we require to prove \prettyref{thm:arbitrary_formal}. We put it all together in the proof below.
\begin{proof}[Proof of \prettyref{thm:arbitrary_formal}]
For $\gamma \geq \ffrac{2}{3}$, we show in \prettyref{prop:farkas_apply} that for range of $p \geq 5\paren{\ffrac{\paren{\log k}}{\gamma^4 k}}^{\ffrac{1}{6}}$ and choice of $t = \paren{\ffrac{56L_{-\tau}}{p^{\ffrac{3}{2}}}}\paren{\sqrt{\ffrac{\log k}{\gamma k}}}$ with high probability (over the randomness of the input), the \prettyref{lp:primal} has a solution. As argued in \prettyref{cor:entrywise_bound} this already implies that $\norm{B}_2 \leq \paren{896/p^{\ffrac{5}{2}}}\sqrt{n \log k}$.  
Given this bound on $\norm{B}_2$, \prettyref{lem:dual_matrix_psd} guarantees that the \prettyref{alg:two} returns the planted set $S$ with high probability (over the randomness of the input).

As argued in \prettyref{lem:handle_adversary}, the solution due to SDP remains optimal even after action of monotone adversary as per \prettyref{step:adversary} in \prettyref{def:semi-random_model}. Therefore \prettyref{alg:two} still return the planted set $S$.
\end{proof}

\section{Exact Recovery using Subspace Enumeration}
In this section, we study the same problem (as in \prettyref{sec:sdp_section}) of exact recovery of a planted bipartite graph, but now in a model constructed according to \prettyref{def:planted_model}. We focus on the regimes when the planted bipartite graph has degree $d = \gamma pk$ and when the planted set has size $k =\Omega_{p}\paren{\sqrt{n}}$. We let $L_{-\tau}$ be the threshold rank of $\restrict{A}{S\times S}$ as defined in \prettyref{def:threshold_rank} for some choice of threshold $\tau$. We let $L^{'}_{-\tau}$ be the threshold rank for the matrix $A$ for some choice of threshold $\tau$.  In \prettyref{sec:partial_recovery}, we give a procedure to recover a list of sets $\mathcal{S}'$ containing a set $S'$ such that $S'$ has $\paren{1-\delta}$ fraction of vertices in $S$ for some constant $\delta >0$. Using arguments similar to \cite{GLR18}, we can recover the remaining set of vertices (\prettyref{lem:matching_argument}). Hence we prove the formal statement of \prettyref{thm:subspace_informal} stated as,

\begin{theorem}[Formal version of \prettyref{thm:subspace_informal}] \label{thm:subspace_formal}
For $n,k,d,p$ satisfying $k \geq \dfrac{768\sqrt{n}}{\gamma p^{3}}$
and for choice of $\tau=(d/2)-2\sqrt{n}$, there exists a deterministic algorithm which can recover the planted set $S$ in an instance generated as per \prettyref{def:planted_model}, exactly with high probability (over the randomness of the input) in time $\bigO\paren{\poly(n){k}^{{{\paren{L_{-\tau}+1}}}}}$.
\end{theorem}

We note that the constants in \prettyref{thm:subspace_formal} have not been optimized for and these specific values are a result of choices we make for ease of calculation.

\subsection{Partial recovery of the planted set} \label{sec:partial_recovery}
In this setting, the vector $\mathbf{u}=\one_{S_1}-\one_{S_2}$, which also indicates the planted set has small Rayleigh quotient (of value $-d$). We recall that this vector $\mathbf{u}$ is referred to as the \textit{signed indicator vector} for our planted set $S$. Although $\mathbf{u}$ is not an eigenvector for the entire matrix $A$, we can still show that it has a large projection on the subspace formed by the bottom $L^{'}_{-\tau'}=L_{-\tau}+1$ eigenvectors (having eigenvalues smaller than $-\tau'$ for $\tau'=\tau+2\sqrt{n}$). Therefore we can do a brute force search in this space via the subspace enumeration technique along the lines of \cite{KT07,ABS10,Kol11,KLT17} and attempt to recover a vector which is close to this signed indicator(distance to the $\mathbf{u}$ is small, see \prettyref{lem:close_vector_in_span}) for the planted set. We then use this vector to recover  $\paren{1-\delta}$ fraction of the planted set $S$  for some constant $\delta >0$.

\begin{lemma} \label{lem:large_fraction_recover}
For an instance of graph generated by \prettyref{def:planted_model} and given a parameter $\delta >0$ and in regimes of $k \geq \paren{(48-6\delta)\sqrt{n}}/\delta \gamma p$, there exists a deterministic algorithm running in time $\bigO\paren{\poly\paren{n}k^{L^{'}_{-\tau'}}}$ that computes a $\bigO\paren{k^{L^{'}_{-\tau'}}}$ sized list of sets $\mathcal{S}'$ such that each list has size $k$ and there exists a set $S' \in \mathcal{S}'$ having $\abs{S \cap S'}\geq \paren{1-\delta}k$.
\end{lemma}

We start with a useful fact about how the eigenvalues of a matrix are shifted after adding another matrix. This will be useful in the analysis later.

\begin{fact}[Weyl's  inequality]
\label{fact:weyl_inequality}
Let $C$ and $B$ be $n \times n$ symmetric matrices with eigenvalues denoted by  $\lambda_1(C),\hdots,\lambda_n(C)$ and $\lambda_1(B),\hdots,\lambda_n(B)$ respectively, then for the eigenvalues of $C+B$ denoted by $\lambda_i(C+B), \forall i \in [n]$ we have that,
\begin{align*}
    \lambda_i(C) +\lambda_1(B) \leq \lambda_i(C+B) \leq \lambda_i(C) + \lambda_n(B) \mper
\end{align*}
\end{fact}

\begin{lemma}\label{lem:relate_threshold_rank}
For $\tau'=\tau+2\sqrt{n}$ we have,
$\rank_{\leq -\tau'} \paren{A} \leq \rank_{\leq -\tau} \paren{\restrict{A}{S\times S}}+1$
\end{lemma}

\begin{proof}
We will relate $\rank_{\leq -\tau}\paren{\restrict{A}{S \times S}}$ to $\rank_{\leq -\tau'}(A)$ for $\tau'=\tau+2\sqrt{n}$. We can express our matrix $A$ in \prettyref{def:planted_model} as sum of \say{simpler} component matrices,
\begin{align}\label{eq:express_random_A}
    A=A_{S \times S} - p\one_S\one_S^T + p\one\one^T + R \mper
\end{align}
We start with the first component i.e $A_{S \times S}$ and note that
\begin{align*}
    \rank_{\leq -\tau}\paren{A_{S \times S}}= \rank_{\leq -\tau}\paren{\restrict{A}{S \times S}}=L_{-\tau} \mper
\end{align*}
Next, we consider the $-p\one_S\one_S^T$ term in equation \prettyref{eq:express_random_A}. This is a rank one matrix and shifts only the eigenvalue corresponding to the $\one_S$ eigenvector of $A_{S \times S}$ from $d$ to $d-pk$. Therefore we have,
\begin{align*}
    \rank_{\leq -\tau}\paren{A_{S \times S}-p\one_S\one_S^T} \leq L_{-\tau}+1 \mper
\end{align*}
Next, we consider the matrix $R$ in equation \prettyref{eq:express_random_A}. Since almost surely $\norm{R}_2 \leq 2\sqrt{n}$ (\prettyref{claim:random_matrix_norm}), and our choice of $\tau'=\tau+2\sqrt{n}$ we therefore have that,
\begin{align*}
    \rank_{\leq -\tau'}\paren{A_{S \times S}-p\one_S\one_S^T+R} 
    &= \abs{i:\lambda_i\paren{A_{S \times S}-p\one_S\one_S^T+R} \leq -\tau'}\\ 
    &\leq \abs{i:\lambda_i\paren{A_{S \times S}-p\one_S\one_S^T}\leq -\tau'+2\sqrt{n}} \\
    &=\abs{i:\lambda_i\paren{A_{S \times S}-p\one_S\one_S^T} \leq -\tau} \leq L_{-\tau}+1 \mper
\end{align*}
Finally we account for the term $p\one\one^T$ in equation \prettyref{eq:express_random_A} by using \prettyref{fact:weyl_inequality} where we set $B=p\one\one^T$ and $C=A_{S\times S}-p\one_S\one_S^T+R$. Here $\lambda_1\paren{B}=0$, and we get that,
\begin{align*}
    \lambda_1\paren{C} \leq \lambda_1\paren{C+B} \leq \lambda_2\paren{C} \leq  \hdots \leq \lambda_{n-1}\paren{C+B} \leq \lambda_n\paren{C} \leq \lambda_n\paren{C+B} \mper 
\end{align*}
Therefore we have that,
\begin{align*}
    \rank_{\leq -\tau'} \paren{A} = \rank_{\leq -\tau'}\paren{C+B} \leq \rank_{\leq -\tau'}\paren{C}  \leq L_{-\tau}+1 \mper
\end{align*}
\end{proof}

\begin{lemma}\label{lem:close_vector_in_span}
Let $\mathbf{v}^{\paren{1}},\hdots,\mathbf{v}^{\paren{n}}$ denote the eigenvectors of $A$, then for $\tau'=d/2$ there exists a vector $\mathbf{y'} \in \mathsf{span}\set{\mathbf{v}^{\paren{1}},\hdots,\mathbf{v}^{\paren{L^{'}_{-\tau'}}}}$ such that,
\begin{align*}
    \norm{\mathbf{y}'-\mathbf{u}}^2 \leq \dfrac{k\sqrt{n}}{(d/2)+3\sqrt{n}} \mper
\end{align*}
\end{lemma}

\begin{proof}
We express the \textit{signed indicator vector} $\mathbf{u}$ in the basis of the eigenvectors of $A$ as,
\begin{align} \label{eq:eigen_basis_express}
    \mathbf{u}=c_1\mathbf{v}^{\paren{1}} + \hdots c_n\mathbf{v}^{\paren{n}} \mper
\end{align}
for some constants $c_1,\hdots,c_n$.
If we consider the vector $\mathbf{y}'=c_1\mathbf{v}^{\paren{1}}+ \hdots c_{L^{'}_{-\tau'}}\mathbf{v}^{\paren{L^{'}_{-\tau'}}}$ for the same constants $c_1,\hdots,c_{L^{'}_{-\tau'}}$ as in equation \prettyref{eq:eigen_basis_express}, by construction it lies in the space spanned by the bottom $L^{'}_{-\tau'}$ eigenvectors of $A$ i.e.,
\begin{align*}
    \mathbf{y}'=c_1\mathbf{v}^{\paren{1}} + \hdots c_{L^{'}_{-\tau'}}\mathbf{v}^{\paren{L^{'}_{-\tau'}}} \text{ \; belongs to \;} \mathsf{span}\set{\mathbf{v}^{\paren{1}},\hdots,\mathbf{v}^{\paren{L^{'}_{-\tau'}}}}\mper
\end{align*}
Now we compute the distance between these vectors and we get,
\begin{align}\label{eq:simple_distance}
    \norm{\mathbf{y}'-\mathbf{u}}^2 = \norm{\sum_{i=L^{'}_{-\tau'}+1}^n{c_i\mathbf{v}_i}}^2 = \sum_{i=L^{'}_{-\tau'}+1}^nc_i^2 \mper
\end{align}
To finish the proof, we need an upper bound on $\sum_{i=L^{'}_{-\tau'}+1}^nc_i^2$. We consider the matrix
$A'=A+(d+3\sqrt{n})I$. We consider the quadratic form of $\mathbf{u}$ with the matrix $A'$ and we get,
\begin{align*}
    \frac{\mathbf{u}^TA'\mathbf{u}}{\mathbf{u}^T\mathbf{u}} = \frac{\mathbf{u}^T{A}\mathbf{u}}{\mathbf{u}^T\mathbf{u}} +  \frac{\mathbf{u}^T\paren{d+3\sqrt{n}I}\mathbf{u}}{\mathbf{u}^T\mathbf{u}} = -d + \paren{d+3\sqrt{n}} = 3\sqrt{n}
\end{align*}
Now we use equation \prettyref{eq:eigen_basis_express} to write the vector $\mathbf{u}$ in expression above as,
\begin{align*}
    3\sqrt{n} 
    &= \dfrac{\mathbf{u}^TA'\mathbf{u}}{\mathbf{u}^T\mathbf{u}} = \dfrac{\paren{\sum_{i}c_i\mathbf{v}^{\paren{i}}}A'\paren{\sum_{i}c_i\mathbf{v}^{\paren{i}}}}{\norm{\mathbf{u}}^2} = \dfrac{\sum_{i=1}^{n}\paren{\lambda_i\paren{A}+d+3\sqrt{n}}c_i^2\norm{\mathbf{v}^{\paren{i}}}^2}{k}\\
    &\geq \dfrac{\paren{\lambda_{{L^{'}_{-\tau'}+1}}\paren{A}+d+3\sqrt{n}}\sum_{i=L^{'}_{{-\tau'}}+1}^{n}{c_i^2}}{k} \numberthis  \label{eq:distance_to_top_space} \mper
\end{align*}
Substituting the bound on $\sum_{i=L^{'}_{-\tau'}+1}^nc_i^2$ from equation \prettyref{eq:distance_to_top_space} in equation \prettyref{eq:simple_distance} and for $\tau'=d/2$ we have,
\begin{align*}
    \norm{\mathbf{y}'-\mathbf{u}}^2   = \sum_{i=L^{'}_{-\tau'}+1}^nc_i^2 \leq \dfrac{3\sqrt{n} k}{\paren{\lambda_{{L^{'}_{-\tau'}+1}}\paren{A}+d+3\sqrt{n}}} \leq \dfrac{3k\sqrt{n}}{(d/2)+3\sqrt{n}} \mper
\end{align*}
\end{proof}

Hence there exists a vector $\mathbf{y}' \in \mathbb{R}^{L^{'}_{-\tau'}}$, which is close to a vector that indicates the set $S$. Next in \prettyref{lem:epsilon_net_construction}, we show how to find a $\mathbf{y}$ that is close to $\mathbf{y}'$, whose existence we have argued in \prettyref{lem:close_vector_in_span}. We do so by a brute force search over the space spanned by these $L^{'}_{-\tau'}$ eigenvectors. We cannot search over the infinite points in the space as such, but we can construct an $\e$-net and choose a value of $\e$ such that we get a point in this space for which the distance to $\mathbf{y}'$ is small enough (smaller than $\varepsilon$ for some carefully chosen $\varepsilon$).

\begin{lemma} \label{lem:epsilon_net_construction}
There exists a deterministic algorithm running in time $\bigO\paren{k^{L^{'}_{-\tau'}}}$ which finds a vector $\mathbf{y} \in \mathsf{span}\set{\mathbf{v}^{\paren{1}},\hdots,\mathbf{v}^{\paren{L^{'}_{-\tau'}}}}$ such that, 
\begin{align*}
    \norm{\mathbf{y}-\mathbf{u}}^2 \leq \dfrac{6k\sqrt{n}}{(d/2)+3\sqrt{n}}
\end{align*}
\end{lemma}

\begin{proof}
We build an $\e$-net such that for any $\mathbf{v} \in \mathsf{span}\set{\mathbf{v}^{\paren{1}},\hdots,\mathbf{v}^{\paren{L^{'}_{-\tau'}}}}$ we have another $\mathbf{v}'$ which belongs to the $\e$-net and is also close to $\mathbf{v}$ such that,
\begin{align*}
    \norm{\mathbf{v}-\mathbf{v}'} \leq \e \mper
\end{align*}
Thus for the vector $\mathbf{y'}$ in \prettyref{lem:close_vector_in_span} we can find a vector $\mathbf{y}$ such that,
\begin{align*}
    \norm{\mathbf{y}-\mathbf{u}}^2 = \norm{\paren{\mathbf{y}- \mathbf{y}'} + \paren{\mathbf{y}'-\mathbf{u}}}^2 \leq 2\paren{\norm{\mathbf{y}-\mathbf{y}'}^2 + \norm{\mathbf{y}'-\mathbf{u}}^2} \leq 2\paren{\e^2 + \dfrac{3\sqrt{n} k}{d/2+3\sqrt{n}}} \mper
\end{align*}
We choose $\varepsilon$ in our $\varepsilon$-net as $\varepsilon = \sqrt{ \ffrac{3k\sqrt{n}}{((d/2)+3\sqrt{n})}}$ so that,
\begin{align*}
    \norm{\mathbf{y}-\mathbf{u}}^2 \leq 2\varepsilon^2 + \dfrac{6k\sqrt{n}}{(d/2)+3\sqrt{n}} = \dfrac{12k\sqrt{n}}{(d/2)+3\sqrt{n}} \mper
\end{align*}

For unit norm vectors, the number of points in an $\varepsilon$-net is upper bounded by $\paren{\ffrac{3}{\e}}^{L^{'}_{-\tau'}}$ (Corollary 4.2.13, \cite{Ver18}). Since our vector $\mathbf{u}$ has squared norm $k$, we consider ball of radius $k$. The volume in the expression in Corollary 4.2.13, \cite{Ver18} is scaled by factor of $k^{L^{'}_-{\tau'}}$. Therefore, we have that the number of points is upper bounded by $\paren{\ffrac{3k   }{\e}}^{L^{'}_{-\tau'}}$
Substituting our choice of $\e^2 = 3\ffrac{\sqrt{n} k}{\paren{d/2+3\sqrt{n}}}$, the number of points (denoted by $\mathcal{N}$) are bounded by,
\begin{align*}
    \mathcal{N} 
    &\leq \paren{\frac{3k}{\varepsilon}}^{L^{'}_{-\tau'}} = \paren{\sqrt{3k}}^{L^{'}_{-\tau'}}\paren{\frac{(d/2)+3\sqrt{n}}{\sqrt{n}}}^{L^{'}_{-\tau'}/2} = \paren{\sqrt{3k}}^{L^{'}_{-\tau'}}\paren{3+\frac{d}{2\sqrt{n}}}^{L^{'}_{-\tau'}/2}\\
    &\leq \paren{3\sqrt{k}}^{L^{'}_{-\tau'}}\paren{1+\frac{k}{6\sqrt{n}}}^{L^{'}_{-\tau'}/2} \leq \paren{3\sqrt{k}}^{L^{'}_{-\tau'}/2}\paren{1+\frac{\sqrt{n}}{6}}^{L^{'}_{-\tau'}/2}\\
    &\leq \paren{3\sqrt{k}}^{L^{'}_{-\tau'}}\paren{\dfrac{k}{9}}^{L^{'}_{-\tau'}/2} \leq \paren{k}^{L^{'}_{-\tau'}} \qquad\quad \paren{\text{Using }3\sqrt{n} \leq k \leq n}
    \mper
\end{align*}
Therefore the number of points are $\bigO\paren{{k}^{L^{'}_{-\tau'}}}$. Hence we can construct this $\e$-net in time $\bigO\paren{{k}^{{{L^{'}_{-\tau'}}}}}$.
\end{proof}

\begin{proof} [Proof of \prettyref{lem:large_fraction_recover}]
For any $0 \leq \delta \leq 1$ and in the regimes of $k \geq \ffrac{\paren{(48-6\delta)\sqrt{n}}}{\delta\gamma p}$, using   \prettyref{lem:close_vector_in_span} we have that,
\begin{align*}
    \norm{\mathbf{y}-\mathbf{u}}^2 \leq \frac{12k\sqrt{n}}{d/2+3\sqrt{n}}= \frac{24k\sqrt{n}}{\gamma pk + 6\sqrt{n}} = \frac{24k}{(\gamma pk/\sqrt{n})+6} \leq \frac{24k}{((48-6\delta)/\delta) + 6} = \frac{\delta k}{2} \mper
\end{align*}

Next we formalize that this vector $\mathbf{y}$ closely indicates our planted set $S$. We sort the entries of the vector $\mathbf{y}$ by absolute value and pick the top $k$ entries in a set ${S}'$. Let $t$ be a threshold such that we have ${S}' = \set{i:\abs{y_i} \geq t}$. We note that $t \leq 1$ since otherwise if $t>1$, for the vector $\mathbf{u}$ which lies inside the $\varepsilon$-net (for our choice of $\varepsilon$), we will have $\norm{\mathbf{u}}^2 >k$. However, we know that this is not true since $\norm{\mathbf{u}}^2 =k$ .

We denote $B$ as the bad set of vertices, $B\defeq S\setminus {S}'$. We let $\eta$ be the fraction of these which belong to $S_1$ and $\paren{1-\eta}$ fraction then belong to $S_2$. Therefore,
\begin{align*}
    \norm{\mathbf{y}-\mathbf{u}}^2 = \sum_{i \notin S}{y_i}^2 + \sum_{i \in S_1}\paren{y_i-1}^2 + \sum_{i \in S_2}\paren{y_i+1}^2 \mper
\end{align*}
Doing a term by term analysis we get that,
\begin{align*}
   \sum_{i \notin S}y_i^2 
    &\geq \sum_{i \in {S}',i \notin S}y_i^2 \geq \abs{B}{t}^2 \\
    \sum_{i \in S_1}{\paren{y_i-1}^2}
    &\geq \sum_{i \in S_1,i \notin {S}'}{\paren{y_i-1}^2} \geq \eta \abs{B} \min\set{(t-1)^2,(t+1)^2} \geq \eta\abs{B}(1-t)^2 \\
    \sum_{i \in S_2}{\paren{y_i+1}^2}
    &\geq \sum_{i \in S_2,i \notin {S}'}\paren{y_i+1}^2 \geq \paren{1-\eta}\abs{B}\min\set{(1-t)^2,(1+t)^2}\\ &\qquad\qquad\qquad\qquad\quad\geq \paren{1-\eta}\abs{B}(1-t)^2 \mper
\end{align*}
where the first inequality holds because $u_i=0$ and $y_i^2 \geq t^2$ and in the second and third inequality we use $u_i=-1$ and $u_i=1$ respectively and $y_i \in [-t,t]$ where $t \leq 1$. Therefore we get that,
\begin{align*}
    \dfrac{\delta k}{2} \geq \norm{\mathbf{y}-\mathbf{u}}^2 \geq \abs{B}\paren{t^2 + (1-t)^2} \geq \dfrac{\abs{B}}{2}\mper
\end{align*}
The last inequality holds by observing that $t=1/2$ minimizes that expression and thus we have that the set of bad vertices $\abs{B} \leq \delta k$.

However, we can't explicitly compute this $\mathbf{y}$. Therefore, we do this thresholding for all vectors in the $\varepsilon$-net (there are $\bigO(k^{L^{'}_{-\tau'}})$ such vectors) from \prettyref{lem:close_vector_in_span}. Thus, we obtain a list (of size $\bigO\paren{k^{L^{'}_{-\tau}}}$) of sets, denoted by $\mathcal{S}'$, such that it contains a set $S'$ of size $k$ where $\abs{S \cap S'}\geq (1-\delta)k$
\end{proof}

\subsection{ Algorithm for full recovery}
In \prettyref{lem:large_fraction_recover} we output a list of sets $\mathcal{S}'$ such that it contains a set $S'$ where $\abs{{S}'}=k$ and $\abs{{S}' \cap S} \geq \paren{1-\delta}k$ for a constant $\delta >0$. In this section, we propose an algorithm that allows us to recover the whole of the planted set (in appropriate parameter regimes). The main idea here is to distinguish between vertices that belong to $S$ and those which don't by considering the size of maximum matching in subgraph induced on the neighborhood of vertex and the set ${S}'$. This idea is used in the work \cite{GLR18} in a similar vein to recover the vertices in a semi-random model for the same problem. 

However, the set ${S}'$ is not a fixed set but a set that is function of the randomness of the input. Therefore, we will bound the size of matching to the fixed set $S$ by using Chernoff bounds.

\begin{lemma}\label{lem:matching_argument}
Given a set ${S}'$ of size $k$ such that $S \cap {S}' \geq \paren{1-\delta}k$ for $\delta \leq p^2/16$ such that $k \geq \ffrac {768\sqrt{n}}{\gamma p^3}$ and $p \geq 5\sqrt{\ffrac{\paren{\log k}}{k}}$, there exists a polynomial time deterministic algorithm that recovers the planted set $S$ completely.
\end{lemma}

\begin{proof}
Let $\mathsf{MM}\paren{v,S}$ denote the size of the maximum matching in the graph induced on $N(v) \cap S$.
For a vertex $v \in V \setminus S$, the expected size of this maximum matching is given as,
\begin{align*}
    \E\Brac{\mathsf{MM}\paren{v,S}} \geq \dfrac{p^2k}{2} \mper
\end{align*}
This is because a $d$-regular bipartite graph has at least one perfect matching. We fix an arbitrary such matching $\mathsf{M}_v$. Each edge in $\mathsf{M}_v$ is present in this graph induced on $N(v) \cap S$ with probability $p^2$. 
In the construction of instance, we picked the edges between $v$ and $N(v)$ independently. Therefore the edges in $N(v) \cap S$ are also independent of each other. Using Chernoff bounds (\prettyref{fact:chernoff_lower}), for a fixed vertex $v \in V \setminus S$  we obtain,
\begin{align*}
    \Pr{ \mathsf{MM}\paren{v,S} \leq \dfrac{p^2k}{4}} \leq \exp\paren{-\dfrac{p^2k}{8}}  \mper
\end{align*}
To make this claim for all vertices in $V \setminus S$ we do a union bound to obtain,
\begin{align*}
    \Pr{\exists v \in V\setminus S, \mathsf{MM}\paren{v,S} \leq \dfrac{p^2k}{4}}  \leq n\exp\paren{-\dfrac{p^2k}{8}} \leq k^2\exp\paren{-\dfrac{p^2k}{8}}  \mper
\end{align*}
Hence, with high probability (over the randomness of the input), for $p \geq 5\sqrt{\ffrac{\paren{\log k}}{k}}$ we have a lower bound on size of matching in the graph induced on the neighbourhood of $v \in V \setminus S$ and the set $S$, for all $v \in V\setminus S$ .
However as we mention earlier, we are interested in size of matching for a vertex $ v \in V\setminus S$ and ${S}'$. Since $\abs{S \setminus S'} \leq \delta k$, and the vertices in matching edges need to be distinct, the size of matching $\mathsf{MM}(v,S')$ drops by at most $\delta k$ when compared to $\mathsf{MM}(v,S)$. Therefore the size of matching for vertex $v \in V \setminus S$ and set ${S}'$ can be lower bounded as,
\begin{align*}
    \mathsf{MM}\paren{v,{S}'} \geq \paren{\dfrac{p^2}{4}-\delta}k \mper
\end{align*}
Now for a vertex $v \in S$ the size of matching in $N(v) \cap S$ is $0$ since $S$ is a bipartite graph and hence it has no triangles. Therefore, using the same argument as above, we can upper bound the size of maximum matching for vertex $v \in S$ in the graph $N(v) \cap {S}'$ as,
\begin{align*}
    \mathsf{MM}\paren{v,{S}'}  \leq \delta k\mper
\end{align*}
Therefore we can distinguish the set to which a  vertex belongs if,
\begin{align} \label{eq:distinguish_vertex}
    \dfrac{ p^2}{4} - \delta > \delta \text{, which is indeed true for $\delta \leq p^2/16$ } \mper
\end{align}
Using this value of $\delta$ in value of $k$ from \prettyref{lem:large_fraction_recover}, we have that $k \geq \ffrac{768\sqrt{n}}{\gamma p^3}$.
\end{proof}

\begin{lemma}\label{lem:unique_planted_bipartite}
For regimes of $p \geq (10\ln n)/\sqrt{k}$ ,the planted bipartite graph $S=(S_1,S_2)$ in \prettyref{def:planted_model} is with high probability (over the randomness of the input), the unique induced bipartite subgraph of size $k$.
\end{lemma}

\begin{proof}
If we consider the graph induced on $V \setminus S$ with $n'=n-k$ vertices, it is a well known (refer \cite{Mat76}) that with high probability (over the randomness of the instance) the size of largest independent set is $\paren{2+o(1)}\log_{1/(1-p)}n'$. which is much smaller than $k/2 = \Omega_p\paren{\sqrt{n}}$. Here we bound the size by,
\begin{align*}
   \paren{2+o(1)}\log_{1/(1-p)}n'+1 \leq 3\log_{1/(1-p)}n=\frac{3\ln n}{-\ln(1-p)} \leq \frac{3 \ln n}{-\ln(e^{-p})}= \frac{3 \ln n}{p} \mper
\end{align*}
Therefore, with high probability (over the randomness of the input), there is no bipartite subgraph of size $k$ which completely lies inside $V \setminus S$.

Also, with high probability (over the randomness of the input), there cannot be a large bipartite graph of size $k$, that spans vertices in both $S$ and $V \setminus S$.
This is because for an induced bipartite subgraph $(R_1,R_2)$ of size $k$, we have $R_1 \cap \paren{V \setminus S} \leq (3\ln n)/p$ and $R_2 \cap \paren{V \setminus S} \leq (3\ln n)/p$.
Fix $S'_1,S'_2$ to be the set of vertices removed from $S_1,S_2$ respectively such that $S'_1=S_1\setminus R_1$ and $S'_2=S_2 \setminus R_2$. We let $T_1,T_2$ be the set of vertices added to $S_1,S_2$ respectively such that $T_1=\paren{V \setminus S_1} \cap R_1$ and $T_2 = \paren{V \setminus S_2} \cap R_2$. Since $\abs{R_1}=\abs{R_2}=k/2$ and $\abs{T_1}\leq (3 \ln n)/p,\abs{T_2}\leq (3\ln n)/p$, we have $\abs{S'_1}\leq (3 \ln n)/p$ and $\abs{S'_2}\leq (3\ln n)/p$ as well. For fixed $S'_1,S'_2,T_1,T_2$, the probability that $(R_1,R_2)$ is a bipartite graph can be bounded as,
\begin{align*}
    \Pr{(R_1,R_2) \text{ is a bipartite graph }} &\leq \Pr{\nexists \text{ edge between $R_1$ and $T_1$}} \times\\
    &\,\,\,\,\,\,\,\Pr{\nexists \text{ edge between $R_2$ and $T_2$}}\\ &\leq \paren{1-p}^{\abs{R_1}\abs{T_1}}\paren{1-p}^{\abs{R_2}\abs{T_2}} =(1-p)^{(k/2)\paren{\abs{T_1}+\abs{T_2}}}\\
    &\leq e^{-(pk/2)\paren{\abs{T_1}+\abs{T_2}}} \leq e^{-pk/2} \quad\paren{\text{Since } \abs{T_1}+\abs{T_2} \geq 1} \mper
\end{align*}
Now using a union bound over all possible $ \binom{n}{\leq (3 \ln n)/p} \leq ((3 \ln n)/p)n^{(3 \ln n)/p}$ 
choices of $S'_1,S'_2,T_1,T_2$ we have that,
\begin{align*}
    \Pr{(R_1,R_2) \text{ is a bipartite graph }}  &\leq e^{-pk/2}n^{(12 \ln n)/p}\paren{\frac{3\ln n}{p}}^4\\
    &= e^{-pk/2}e^{\paren{(12 \ln n)/p} \ln n}\paren{\frac{3\ln n}{p}}^4\\
    &=e^{-pk/2}e^{\paren{(12 \ln n)/p} \ln n}e^{4 \ln ((3 \ln n)/p)}\\
    &\leq e^{-pk/2}e^{(12\ln^2 n)/p}e^{4 \ln ((3 \ln n)/p)}
    \mper
\end{align*}
For $p \geq (10\ln n)/\sqrt{k}$, we bound the $e^{(12 \ln^2 n)/p}$ term as $e^{(12 \ln^2 n)/p} \leq e^{12p^2k/100p} \leq e^{pk/8}$. The term $e^{4 \ln ((3 \ln n)/p)}$ in the expression above can be bound using the inequality $\ln x \leq x$ and the bound we just obtained on the term $e^{(12 \ln^2 n)/p}$ as,
\begin{align*}
    e^{4 \ln ((3 \ln n)/p)} \leq e^{(12\ln n)/p} \leq e^{(12\ln^2 n)/p} \leq e^{pk/8} \mper
\end{align*}
Now, we note that $p \geq (10 \ln n)/\sqrt{k}$ implies that $p \geq (8\ln k)/k$ and,
\begin{align*}
    \Pr{(R_1,R_2) \text{ is a bipartite graph }}  &\leq e^{-pk/2}e^{pk/8}e^{pk/8} = e^{-pk/4} \leq \frac{1}{k^2} \leq \frac{1}{n}
    \mper
\end{align*}

Therefore, with high probability (over the randomness of the instance), the planted set $S$ is the unique induced bipartite subgraph of size $k$.
\end{proof}

\begin{proof}[Proof of \prettyref{thm:subspace_formal}]
For $k \geq 768\sqrt{n}/\gamma p^3$, if we choose $\delta \leq p^2/16$, we have that $k \geq \paren{(48-6\delta)\sqrt{n}}/\delta \gamma p$ and application of \prettyref{lem:large_fraction_recover} returns a $\bigO\paren{k^{L^{'}_{-\tau'}}}$ sized list of sets $\mathcal{S}'$ such that it contains a set $S'$ where $\abs{S \cap S'}\geq \paren{1-\delta}k$. For the regimes of $k \geq 768\sqrt{n}/\gamma p^3$ and $p \geq 5\sqrt{\log k/k}$, \prettyref{lem:matching_argument} allows us to compute the exact planted set $S$ from $S'$ (with high probability over the randomness of the input). We can check in polynomial time that the set returned in \prettyref{lem:matching_argument} is indeed a bipartite graph. 

For other sets in the list $\mathcal{S}'$, we will apply the same steps (from \prettyref{lem:large_fraction_recover} and \prettyref{lem:matching_argument}).
However, for $p \geq (10\ln n)/\sqrt{k}$, \prettyref{lem:unique_planted_bipartite} argues that with high probability (over the randomness of the input), the planted graph $S$ is the unique bipartite graph of size $k$. We note that $p \geq \max\set{5\sqrt{\log k/k},(10\ln n)/\sqrt{k}} = (10\ln n)/\sqrt{k}$, and $k \geq 768\sqrt{n}/\gamma p^3$ implies that $p \geq (10\ln n)/\sqrt{k}$. 
Therefore, for $k \geq 768\sqrt{n}/\gamma p^3$ with high probability (over the randomness of the input), we can recover the planted set $S$ exactly.

Now, iterating over the list $\mathcal{S}'$, computing the eigenvectors and steps in \prettyref{lem:close_vector_in_span} and  \prettyref{lem:matching_argument} take $\bigO\paren{\poly(n)k^{L^{'}_{-\tau'}}}$ time. Therefore, the overall running time is $\bigO\paren{\poly(n)k^{L^{'}_{-\tau'}}}$. Using \prettyref{lem:relate_threshold_rank} the running time is $\bigO\paren{\poly(n)k^{L_{-\tau}+1}}$
\end{proof}

\paragraph{Acknowledgements.}
AK received funding from the European Research Council (ERC) under the European Union’s Horizon 2020 research and innovation programme (grant
agreement No 759471). 
AL was supported in part by SERB Award ECR/2017/003296 and a Pratiksha Trust Young Investigator Award. 

\bibliographystyle{amsalpha}
\bibliography{references}

\appendix
\section{Omitted proofs}

\subsection{Computing the dual of \prettyref{sdp:primal}}
\label{app:lagrangian_calculation}
We compute the dual by writing SDP in matrix form as,
\begin{mybox}
		\begin{SDP}
		\label{sdp:aprimal}
		\[ \min \inprod{A,X}  \]
		\subjectto
		\begin{align}
			&\inprod{I,X} = 1\\
			&\inprod{\one_{ij},X} \leq 0 & \forall \set{i,j} \in E\\
			&\inprod{\one_{ji},X} \leq 0 & \forall \set{i,j} \in E\\
			&k\inprod{D_i,X} \leq 1 & \forall i \in V\\
			&X \succeq 0\mper
		\end{align}
		\end{SDP}
	\end{mybox}

\noindent	
and we compute the Lagrangian associated with \prettyref{sdp:aprimal} as,
\begin{align*}
    \hspace{-1em}
    \mathcal{L}\paren{X,B,Y,\beta} 
    &= \inprod{A,X} + \beta(1-\inprod{I,X}) + \sum_{i \in V}\gamma_i\paren{k\inprod{D_i,X}-1}\\
    &+ \sum_{\set{i,j}\in E}{B_{ij}\inprod{\one_{ij},X}} + \sum_{\set{i,j}\in E}{B_{ji}\inprod{\one_{ji},X}} - \inprod{Y,X}\\
    &=\inprod{A-\beta I+k\sum_{i \in V}\gamma_iD_i+B-Y,X} + \beta-\sum_{i \in V}\gamma_i
\end{align*}
where $\beta$ is unconstrained Lagrange dual variable for the equality constraint and  $B$ is the matrix of non-negative Lagrange variables $B_{ij}$'s and $B_{ji}$'s for the inequality constraints and $Y$ is a p.s.d Lagrange dual matrix variable . We then compute the corresponding Lagrange dual function as,
\begin{align*}
    g\paren{Y,B,\beta} = \inf_{X \succeq 0}{\mathcal{L}\paren{X,B,Y,\beta}} = 
    \begin{cases}
    \beta - \sum\limits_{i \in V}\gamma_i \text{ \quad\quad if } A-\beta I + k\sum\limits_{i \in V}\gamma_iD_i + B - Y = 0\\
    -\infty \text{ \,\,\,\quad\quad\qquad otherwise}
    \end{cases}
\end{align*}
Therefore we obtain the dual SDP program as described in \prettyref{sdp:dual}

\subsection{Proof of \prettyref{fact:farkas_variant} }
\begin{proof}\label{app:farkas_variant_proof}
The first system of equations $\set{\mathbf{x}:C\mathbf{x}=\mathbf{f}, \mathbf{0} \leq \mathbf{x} \leq \mathbf{u}}$ can equivalently be written using slack variables $\mathbf{x}'$ as the system of equations  $\set{C\mathbf{x}=\mathbf{f}, \mathbf{x}+\mathbf{x}'=\mathbf{u},\mathbf{x} \geq 0,\mathbf{x}' \geq 0}$. We can rewrite the latter system of equations in the standard $\set{\tilde C\tilde{\mathbf{x}}= \tilde{\mathbf{f}},\tilde{\mathbf{x}}\geq 0}$ form by writing the system of equations as,
\begin{align*}
    \begin{bmatrix}
    C & 0\\
    I & I
    \end{bmatrix}
    \begin{bmatrix}
    \mathbf{x}\\
    \mathbf{x}'
    \end{bmatrix}= 
    \begin{bmatrix}
    \mathbf{f}\\
    \mathbf{u}
    \end{bmatrix}
\end{align*}
where the matrix is $\tilde{C}$ and $\tilde{\mathbf{x}} = \begin{bmatrix}
\mathbf{x} &\mathbf{x}'
\end{bmatrix}^T$ and $\tilde{\mathbf{f}} = \begin{bmatrix}
\mathbf{f} &\mathbf{u}
\end{bmatrix}^T$.
Now using standard variant of Farkas' Lemma the corresponding dual system is $\set{\tilde{C}^T\tilde{\mathbf{y}} \geq 0, \tilde{\mathbf{f}}^T\tilde{\mathbf{y}}<0}$ where $\tilde{\mathbf{y}} = \begin{bmatrix}
\mathbf{y} &\mathbf{z}
\end{bmatrix}^T$.
We can rewrite the constraint $\tilde{C}^T\tilde{\mathbf{y}}^T \geq 0$ of the dual system as,
\begin{align*}
    \begin{bmatrix}
    C & 0\\
    I & I
    \end{bmatrix}^T
    \begin{bmatrix}
    \mathbf{y}\\
    \mathbf{z}
    \end{bmatrix} =
    \begin{bmatrix}
    C^T & I\\
    0 & I
    \end{bmatrix}
    \begin{bmatrix}
    \mathbf{y}\\
    \mathbf{z}
    \end{bmatrix} =
    \begin{bmatrix}
    {C^T\mathbf{y}+\mathbf{z}}\\
    \mathbf{z}
    \end{bmatrix}
    \geq 0
\end{align*}
and we get the constraints that $\mathbf{z} \geq 0$ and $C^T\mathbf{y}+\mathbf{z} \geq 0$ as in \prettyref{fact:farkas_variant}. Next, we consider the dual constraint $\tilde{\mathbf{f}}^T\tilde{\mathbf{y}}<0$  and we can rewrite it as
\begin{align*}
    \begin{bmatrix}
    \mathbf{f}&
    \mathbf{u}
    \end{bmatrix}
    \begin{bmatrix}
    \mathbf{y}\\
    \mathbf{z}
    \end{bmatrix} <0
\end{align*}
which gives the constraint $\mathbf{f}^T\mathbf{y}+ \mathbf{u}^T\mathbf{z}<0$ from \prettyref{fact:farkas_variant}.
Therefore, we obtain the second system of equations 
$\set{\mathbf{y}:C^T\mathbf{y} + \mathbf{z}  \geq 0,\mathbf{f}^T\mathbf{y} + \mathbf{u}^T\mathbf{z}<0,\mathbf{y} \in \mathbb{R}^{L_{-\tau}},\mathbf{z} \geq 0} \mper$

Now, that we show that these are dual systems, using the standard variant of Farkas' Lemma (refer Chapter 5,\cite{BV04}) we have that only one of these system of equations has a solution. 
\end{proof}

\subsection{Proof of \prettyref{fact:optimality_conditions}}
\label{app:sdp_complementary_slackness}
\begin{proof}
First we argue that $X$ has to be a rank one matrix. We consider a proof by contradiction and let $X = \sum\limits_{i=1}^{n}\lambda_i \mathbf{v}_i\mathbf{v}_i^T$ where $\lambda_i \geq 0$ (since $X \succeq 0$) and $\mathbf{v}_{i}$'s are orthonormal. By complementary slackness we have,
\begin{align}
    X\cdot Y = 0 \implies \sum_{i=1}^{n} \lambda_i \mathbf{v}_i^TY\mathbf{v}_i = 0 \mper
\end{align}
Since $Y \succeq 0$, we have $\mathbf{v}_i^TY\mathbf{v}_i \geq 0,\lambda_i \geq 0, \forall i \in [n]$ and hence $\lambda_i\mathbf{v}_i^TY\mathbf{v}_i=0,\forall i \in [n]$.
Since $Y \succeq 0$, we have $Y=C^TC$ for some C and whenever $\lambda_i \neq 0$, it implies that $\mathbf{v}_i^TY\mathbf{v}_i = \mathbf{v}_i^TC^TC\mathbf{v}_i = \inprod{C\mathbf{v}_i,C\mathbf{v}_i}=\norm{C\mathbf{v}_i}^2=0$ and hence $C\mathbf{v}_i=0$. Now, $Y\mathbf{v}_i =C^T(C\mathbf{v}_i)= 0$ and hence $\mathbf{v}_i \in \mathsf{nullity}(Y)$. Now for sake of contradiction we consider $\mathsf{rank}(X) \geq 2$, then we have some $\lambda_i,\lambda_j \neq 0$ and therefore $\mathbf{v}_i,\mathbf{v}_j \in \mathsf{nullspace}(Y)$. Since $\mathbf{v}^{(i)},\mathbf{v}^{(j)}$ are orthonormal we have $\mathsf{nullity}(Y) \geq 2$. Using rank-nullity theorem we get $\mathsf{rank}(Y) = n- \mathsf{nullity}(Y) \leq n-2$.
This contradicts the fact that $\mathsf{rank}(Y)=n-1$.

Also, $X=\mathbf{g}\mathbf{g}^T$ is the unique solution since otherwise if there is some other rank one solution (say $X=\mathbf{h}\mathbf{h}^T$), any convex combination of these has the same objective value and gives us an optimal (in objective function value) rank two solution.
\end{proof}

\subsection{Proof of \prettyref{fact:matrix_expt}}
\begin{proof} \label{app:matrix_expt}
By our definition of $M$ we have that,
\begin{align*}
    M_{rr}= \sum_{i \in N(j)}\paren{{\mathbf{w}^{\paren{i}}{\mathbf{w}^{\paren{i}}}^T}}_{rr} =  \sum_{i \in N(j)}\paren{w^{\paren{i}}_r}^2 
\end{align*}
and taking expectation over the choice of random edges $i \in N(j)$ we obtain,
\begin{align*}
    \E\Brac{M_{rr}} = \E\Brac{ \sum_{i \in S}\one_{i \in N(j)}\paren{w^{\paren{i}}_r}^2} = \sum_{i \in S}{\E\Brac{\one_{i \in N(j)}\paren{w^{\paren{i}}_r}^2}}
\end{align*}
where $\one_{i \in N(j)}$ is an indicator random variable for the edge $\set{i,j}$ and takes the value $1$ with probability $p$ and $0$ with probability $1-p$. Therefore,
\begin{align*}
    \E\Brac{M_{rr}} = \sum_{i \in S}{\E\Brac{\one_{i \in N(j)}\paren{w^{\paren{i}}_r}^2}} = p\sum_{i \in S}{\paren{w^{\paren{i}}_r}^2} = p\sum_{i \in S}{\paren{v^{\paren{r}}_i}^2} = p\norm{\mathbf{v}^{\paren{r}}}^2 = p
    \mper
\end{align*}
Similarly we obtain $\E\Brac{M_{rs}}$ for $r\neq s$ as
\begin{align*}
    \E\Brac{M_{rs}} 
    &= \E\Brac{\sum_{i \in N(j)}w^{\paren{i}}_rw^{\paren{i}}_s} = \E\Brac{\sum_{i \in S}\one_{i \in N(j)}w^{\paren{i}}_rw^{\paren{i}}_s} = \sum_{i \in S}\E\Brac{\one_{i \in N(j)}w^{\paren{i}}_rw^{\paren{i}}_s}\\ 
    &= p\sum_{i \in S}w^{\paren{i}}_rw^{\paren{i}}_s = p\sum_{i \in S}v^{\paren{r}}_iv^{\paren{s}}_i = p\inprod{\mathbf{v}^{\paren{r}},\mathbf{v}^{\paren{s}}}=0
    \mper
\end{align*}
\end{proof}

\section{Standard concentration inequalities}

\begin{fact}[Hoeffding Bound, Theorem 4.14 - \cite{MU18}]
\label{fact:hoeffding_bound}
Let $X_1,\hdots,X_n$ be independent random variables with $\E\Brac{X_i}=\mu_i$ and $\Pr{a_i \leq X_i \leq b_i}=1$ for constants $a_i$ and $b_i$. Then,
\begin{align*}
    \Pr{\abs{\sum_{i=1}^n{X_i} - \sum_{i=1}^n{\mu_i}}\geq \e} \leq 2\exp\paren{\dfrac{-2\e^2}{\sum_{i=1}^n{\paren{b_i-a_i}^2}}}
\end{align*}
\end{fact}

\begin{fact}[Chernoff bound (Multiplicative); Theorem 4.5 (Part 2) - \cite{MU18}]
	\label{fact:chernoff_lower}
	Let $X_1, X_2, \hdots, X_n$ be i.i.d. bernoulli variables and $X=\sum_{i=1}^nX_i$ such that $\mu = \E[X]$. Then for any $\delta \in (0, 1)$,
	\[
	\ProbOp\left[{\sum_{i=1}^{n} X_i} \leq  (1-\delta) \mu \right] \leq \exp\paren{-\dfrac{\mu\delta^2}{2}}\mper
	\]
\end{fact}

\begin{fact}[Chernoff bound (Multiplicative); Theorem 4.4 (Part 2) - \cite{MU18}]
	\label{fact:chernoff_upper}
	Let $X_1, X_2, \hdots, X_n$ be i.i.d. bernoulli variables and $X=\sum_{i=1}^nX_i$ such that $\mu = \E[X]$. Then for any $\delta \in (0, 1)$,
	\[
	\ProbOp\left[{\sum_{i=1}^{n} X_i} \geq  (1+\delta) \mu \right] \leq \exp\paren{-\dfrac{\mu\delta^2}{3}}\mper
	\]
\end{fact}

\end{document}